\documentclass[english,aps,superscriptaddress,preprint]{revtex4-1}
\usepackage[T1]{fontenc}
\usepackage[latin9]{inputenc}
\usepackage{color}
\usepackage{babel}
\usepackage{verbatim}
\usepackage{amstext}
\usepackage{amssymb}
\usepackage{graphicx}
\usepackage{esint}
\usepackage[unicode=true,pdfusetitle,
 bookmarks=true,bookmarksnumbered=false,bookmarksopen=false,
 breaklinks=false,pdfborder={0 0 0},pdfborderstyle={},backref=false,colorlinks=true]
 {hyperref}
\hypersetup{
 linkcolor=blue, citecolor=cyan}

\makeatletter
\usepackage{xcolor}

\makeatother

\begin{document}
\title{Electric and magnetic conductivities in magnetized fermion systems}
\author{Hao-Hao Peng}
\affiliation{Department of Modern Physics, University of Science and Technology
of China, Hefei 230026, China}
\author{Xin-Li Sheng}
\affiliation{INFN Sezione di Firenze, Via Giovanni Sansone, 1, 50019 Sesto Fiorentino}
\author{Shi Pu}
\affiliation{Department of Modern Physics, University of Science and Technology
of China, Hefei 230026, China}
\author{Qun Wang}
\affiliation{Department of Modern Physics, University of Science and Technology
of China, Hefei 230026, China}
\affiliation{School of Mechanics and Physics, Anhui University of Science and Technology,
Huainan 232001, China}
\begin{abstract}
In Wigner function approach with relaxation time approximation, we
calculate electric and magnetic conductivities of a fermion system
in the strong magnetic field. The linear response has been calculated
to the perturbation of electromagnetic fields on the background constant
magnetic field. The Wigner function is separated into an equilibrium
part in the background magnetic field and an off-equilibrium part
induced by perturbative fields. The analytical expression for the
equilibrium part and the corresponding equilibrium conditions are
given. For the off-equilibrium part, we obtain the kinetic equation
at the leading order in $\hbar$ from the master equation of the Wigner
function. When perturbative fields only depend on the proper time,
the off-equilibrium part can be analytically solved from which the
vector and axial vector currents are obtained. We obtain the longitudinal
and transverse Ohm conductivities as well as Hall conductivity as
the linear response of the vector current to the perturbative electric
field. The behaviors of these conductivities as functions of the evolving
time, relaxation time, particle mass, and strength of the background
magnetic field are investigated both analytically and numerically.
\end{abstract}
\maketitle

\section{Introduction}

Relativistic heavy ion collisions provide a unique opportunity to
study strong interaction matter, namely the quark gluon plasma (QGP),
under extreme conditions. Strong electromagnetic (EM) fields are created
in high energy collisions. In peripheral collisions, the dominant
component is the magnetic field along the direction of the reaction
plane \citep{Skokov:2009qp,Voronyuk:2011jd,Bzdak:2011yy,Deng:2012pc}.
In Au+Au collisions at $\sqrt{s_{\text{NN}}}=200$ GeV, the highest
collision energy at the Relativistic Heavy Ion Collider (RHIC) of
Brookhaven National Laboratory, the magnetic field at the moment after
the collision can reach $10^{17}$-$10^{18}$ Gauss \citep{Skokov:2009qp,Voronyuk:2011jd,Bzdak:2011yy,Deng:2012pc,Roy:2015coa}.
In Pb+Pb collisions with $\sqrt{s_{\text{NN}}}=2.76$ TeV at the Large
Hadron Collider (LHC) of the European Organization for Nuclear Research
(CERN), the magnetic field can reach 10 times as high as that at RHIC.
In the early stage of the collision, the magnetic field decays with
time as $\sim t^{-3}$. However, the finite Ohm conductivity of the
QGP slows down the damping of the magnetic field in the later stage,
resulting in a $\sim t^{-1}$ behavior \citep{Tuchin:2013apa,Gursoy:2014aka,Zakharov:2014dia,Yan:2021zjc,Wang:2021oqq,Zhang:2022lje,Siddique:2019gqh,Wang:2020qpx,Pu:2016ayh,Pu:2016bxy,Roy:2015kma}.
This may significantly extend the lifetime of the magnetic field,
so one can expect sizeable effects from the interaction between the
magnetic field and the QGP. These effects include the charge-odd directed
flow $v_{1}^{\text{odd}}$ induced by the Lorentz force acting on
charged particles \citep{Gursoy:2014aka,Inghirami:2019mkc,Gursoy:2020jso,Dubla:2020bdz,Zhang:2022lje},
the splitting of $\Lambda$ and $\overline{\Lambda}$ hyperons' global
polarizations induced by the coupling between the magnetic field and
hyperon's magnetic moment \citep{Muller:2018ibh,Guo:2019joy,Buzzegoli:2022qrr,Xu:2022hql,Peng:2022cya},
and anomalous transport phenomena including the chiral magnetic effect
(CME) \citep{Vilenkin:1980fu,Kharzeev:2007tn,Kharzeev:2007jp,Fukushima:2008xe,STAR:2021mii},
the chiral separation effect (CSE) \citep{Aghababaie:2003iw,Son:2004tq,Metlitski:2005pr},
and the chiral magnetic wave (CMW) \citep{Kharzeev:2010gd,Burnier:2011bf},
see Refs. \citep{Miransky:2015ava,Huang:2015oca,Kharzeev:2015znc,Hattori:2016emy,Fukushima:2018grm,Hattori:2022hyo,Gao:2020vbh,Hidaka:2022dmn}
for reviews. Study of strong interaction matter in strong magnetic
fields may also help us understand compact stars such as magnetars
\citep{Duncan:1992hi,Lai:2000at,Price:2006fi,Enoto:2019vcg}, in which
strong magnetc fields and QGP coexist in the cores of compact stars.


In strong magnetic fields, quark spectrum will be described by Landau
energy levels \citep{landau2013quantum}. As a result, the transport
properties of the QGP will be affected when the field strength is
much larger than the energy scale of the system such as the temperature
or chemical potential. In past few years, many studies have been made
on magnetic field effects on transport quantities such as shear and
bulk viscosities \citep{Hattori:2017qih,Li:2017tgi,Kurian:2018dbn,Kurian:2018qwb,Denicol:2018rbw,Denicol:2019iyh},
diffusion constants \citep{Mitra:2016zdw,Hattori:2017qih,Kurian:2018qwb,Denicol:2018rbw,Denicol:2019iyh,Rath:2019vvi},
jet quenching parameters \citep{Li:2016bbh,Finazzo:2016mhm,Banerjee:2021sjm},
and electrical conductivities \citep{Hattori:2016cnt,Hattori:2016lqx,Gorbar:2016qfh,Fukushima:2017lvb,Kurian:2018qwb,Rath:2019vvi,Lin:2019fqo}.
There are also some effects of strong magnetic fields on photon emission
in heavy-ion collisions \citep{Wang:2022jxx,Wang:2021ebh,Wang:2020dsr,Sun:2023pil}.
Among all these quantities, electrical conductivities are of special
importance because they directly control the lifetime of magnetic
fields and are also necessary parameters in the numerical simulation
of magnetohydrodynamics \citep{anile2005relativistic,giulini2015luciano}.


When the strength of the magnetic field is sufficiently large, the
energy gap between the lowest Landau level and higher Landau levels
becomes so large that nearly all fermions occupy the lowest Landau
level. As a consequence, the system is greatly simplified to a (1+1)-dimension
system from a (3+1)-dimension system, as the transverse motion (perpendicular
to the direction of magnetic field) of fermions is restricted \citep{landau2013quantum}.
Electrical conductivities in the approximation of the lowest Landau
level have been calculated in Ref. \citep{Hattori:2016cnt,Hattori:2016lqx}
with quantum field theory at finite temperature, and in Ref. \citep{Lin:2019fqo}
with chiral kinetic theory. On the other hand, conductivities in weak
magnetic fields have been studied in the kinetic approach \citep{Rath:2019vvi,Gorbar:2016qfh},
holograph models \citep{Pu:2014fva,Pu:2014cwa}, and the Lattice QCD
\citep{Astrakhantsev:2019zkr,Buividovich:2010tn}. In the medium magnetic
field, one has to sum over all Landau levels. Electrical conductivities
in this region have been computed by solving the linearized Boltzmann
equations with explicit $1\leftrightarrow2$ processes in Ref. \citep{Fukushima:2017lvb}
and by employing the relaxation time approximation \citep{Das:2019pqd}
and the holographic QCD model \citep{Li:2018ufq,Fukushima:2021got}.


The present work is based on covariant Wigner functions \citep{Elze:1986qd,Vasak:1987um,Heinz:1983nx,Mrowczynski:1992hq,Blaizot:2001nr,Wang:2001dm,Gao:2012ix,Hidaka:2016yjf,Weickgenannt:2019dks,Sheng:2021kfc},
for recent reviews, see e.g. Refs. \citep{Gao:2020pfu,Hidaka:2022dmn}.
In the constant magnetic field, the Wigner function can be analytically
derived using the exact solution to the Dirac equation \citep{Sheng:2017lfu,Gorbar:2017awz},
which can serve as a tool to calculate quantities such as the pair
production \citep{Sheng:2018jwf} and chiral susceptibility \citep{Das:2019crc}.
For particle collisions or interaction among particles in the Wigner
function approach, we apply the relaxation time approximation as in
Ref. \citep{hakim1992relaxation,Wang:2021qnt}.


This paper is organized as follows. In Sec. \ref{sec:Wigner-function},
we give an introduction to covariant Wigner functions and kinetic
equations under the relaxation time approximation. In Sec. \ref{sec:Equilibrium-Wigner-function},
we discuss Wigner functions in equilibrium to give analytical solutions
and equilibrium conditions. Section \ref{sec:off-equilibrium Wigner function}
is devoted to the kinetic equation and solution for the off-equilibrium
part. One can show that the off-equilibrium part of the Wigner function
is fully determined by its vector and axial vector components. We
focus on these two components in Sec. \ref{sec:Vector-current-induced}
and Sec. \ref{sec:Axial-vector-current-induced}, respectively. Analytical
solutions are given when perturbative fields only depend on the proper
time. The Ohm and Hall conductivities are studied in Sec. \ref{sec:Vector-current-induced}
in respect to their dependence on the evolution and relaxation time,
the particle mass, and the strength of the background magnetic field.
We make conclusions in the last section, Sec. \ref{sec:Conclusion}.


Throughout this work, we choose the convention for the metric tensor
$g_{\mu\nu}=\mathrm{diag}\{+,-,-,-\}$ and the Levi-Civita tensor
$\epsilon^{\mu\nu\rho\sigma}$, $\epsilon^{0123}=-\epsilon_{0123}=+1$.
The local collective velocity $u^{\mu}=\gamma(1,\mathbf{v})$ satisfies
$u^{\mu}u_{\mu}=1$ with $\gamma$ being the Lorentz contraction factor.
Relative to $u^{\mu}$, we define the orthogonal projector $\Delta^{\mu\nu}=g^{\mu\nu}-u^{\mu}u^{\nu}$
. The electric charge $q$ is set to 1 throughout this paper for simplicity.


In this paper, we consider a weak perturbation field $f^{\mu\nu}$
in additional to a strong background field $F_{0}^{\mu\nu}$. The
background field is taken as a constant magnetic field in the comoving
frame with respect to the velocity $u^{\mu}$, which is given by 
\begin{equation}
F_{0}^{\mu\nu}=B_{0}\epsilon^{\mu\nu\alpha\beta}u_{\alpha}b_{\beta},\label{eq:background field}
\end{equation}
where $B_{0}$ is the background field's strength with $b^{\mu}$
being its direction which is orthogonal to $u^{\mu}$, $u_{\mu}b^{\mu}=0$.
We define another projection tensor 
\begin{equation}
\Xi^{\mu\nu}\equiv g^{\mu\nu}-u^{\mu}u^{\nu}+b^{\mu}b^{\nu}.\label{eq:projection operator}
\end{equation}
which can project a four-vector to the direction perpendicular to
both $u^{\mu}$ and $b^{\mu}$.


\section{Covariant Wigner functions \label{sec:Wigner-function}}

The Wigner function is defined as the Fourier transformation of the
two-point correlation function \citep{Elze:1986qd,Vasak:1987um} 
\begin{equation}
W_{\alpha\beta}(x,p)=\int\frac{d^{4}y}{(2\pi\hbar)^{4}}\exp\left(-\frac{i}{\hbar}p\cdot y\right)\left\langle :\overline{\psi}_{\beta}(x_{1})U(x_{1},x_{2})\psi_{\alpha}(x_{2}):\right\rangle ,\label{eq:wf-def}
\end{equation}
where $\left\langle :O:\right\rangle $ denotes the expectation value
of a normal-ordered operator $O$. Here $x_{1}$ and $x_{2}$ are
two space-time coordinates, with $y^{\mu}=x_{1}^{\mu}-x_{2}^{\mu}$
and $x^{\mu}=(x_{1}^{\mu}+x_{2}^{\mu})/2$ being the relative and
central coordinate respectively. With the vector potential $\mathbb{A}_{\mu}$
for the classical electromagnetic field, the gauge link in Eq. (\ref{eq:wf-def})
is defined as 
\begin{eqnarray}
U(x_{1},x_{2}) & = & \exp\left[-\frac{i}{\hbar}y^{\mu}\int_{-1/2}^{1/2}dt\mathbb{A}_{\mu}(x+ty)\right].
\end{eqnarray}
The free-streaming Dirac field satisfies the Dirac equation and its
adjoint form 
\begin{equation}
(i\hbar\gamma\cdot D-m)\psi=\overline{\psi}(i\hbar\gamma\cdot D^{\dagger}+m)=0,
\end{equation}
where $D_{\mu}=\partial_{x\mu}+(i/\hbar)\mathbb{A}_{\mu}$ is the
covariant derivative. The kinetic equation for the Wigner function
has a similar form
\begin{equation}
(\gamma\cdot K-m)W(x,p)=0,\label{eq:kinetic equation}
\end{equation}
which can be derived from the Dirac equation. The operator $K^{\mu}$
is given by 
\begin{eqnarray}
K^{\mu} & = & \Pi^{\mu}+\frac{i\hbar}{2}\nabla^{\mu},\label{eq:operator Kmu}
\end{eqnarray}
where the generalized momentum operator $\Pi^{\mu}$ and space-time
derivative $\nabla^{\mu}$ are 
\begin{eqnarray}
\Pi^{\mu} & = & p^{\mu}-\frac{\hbar}{2}j_{1}(\Delta)F^{\mu\nu}(x)\partial_{p\nu},\nonumber \\
\nabla^{\mu} & = & \partial_{x}^{\mu}-j_{0}(\Delta)F^{\mu\nu}(x)\partial_{p\nu},
\end{eqnarray}
with $j_{0}(\Delta)$ and $j_{1}(\Delta)$ are spherical Bessel functions
and $\Delta=(\hbar/2)\partial_{p}^{\nu}\partial_{x\nu}$. We note
that the derivative $\partial_{x\nu}$ in $\Delta$ only acts on $F^{\mu\nu}(x)$
but not on the Wigner function.


Concerning interaction among particles, we use the relaxation time
approximation in addition to Eq. (\ref{eq:kinetic equation}) \citep{hakim1992relaxation,Wang:2021qnt},
\begin{equation}
(\gamma\cdot K-m)W(x,p)=-\frac{i\hbar}{2}\gamma^{\mu}u_{\mu}\frac{\delta W(x,p)}{\tau},\label{eq:relaxation_time_equation}
\end{equation}
where $\tau$ is the relaxation time, $\delta W$ represents the deviation
from thermodynamical equilibrium
\begin{equation}
\delta W(x,p)\equiv W(x,p)-W_{\text{eq}}(x,p),
\end{equation}
and $u_{\mu}$ is the local collective velocity 
\begin{equation}
u_{\mu}\propto\int d^{4}p\,\text{Tr}\left[\gamma^{\mu}W_{\text{eq}}(x,p)\right].
\end{equation}
In this work, the equilibrium Wigner function is taken as the solution
in a constant $F_{0}^{\mu\nu}$ to the following equation
\begin{equation}
\left\{ \gamma_{\mu}\left[p^{\mu}+\frac{i\hbar}{2}\left(\partial_{x}^{\mu}-F_{0}^{\mu\nu}\partial_{\nu}^{p}\right)\right]-m\right\} W_{\text{eq}}(x,p)=0.\label{eq:equilibrium equation}
\end{equation}
Then from Eq. (\ref{eq:relaxation_time_equation}) we obtain the equation
for $\delta W$ %
\begin{equation}
\left\{ \gamma_{\mu}p^{\mu}+\frac{i\hbar}{2}\gamma_{\mu}\left(\partial_{x}^{\mu}-F_{0}^{\mu\nu}\partial_{\nu}^{p}+\frac{1}{\tau}u^{\mu}\right)-m\right\} \delta W(x,p)=\frac{i\hbar}{2}\gamma_{\mu}f^{\mu\nu}\partial_{\nu}^{p}W_{\text{eq}}+\mathcal{O}(\hbar^{2})\:,\label{eq:equation_for_deviation}
\end{equation}
where we have made a semi-classical expansion with respect to $\hbar$
and truncate it at second order in $\hbar$.


\section{Wigner functions in constant magnetic field: equilibrium \label{sec:Equilibrium-Wigner-function}}

In this section, we discuss about the equilibrium form of the Wigner
function $W_{\text{eq}}(x,p)$ as the solution to Eq. (\ref{eq:equilibrium equation})
in the constant magnetic field (\ref{eq:background field}) without
interaction among particles.

The Wigner function satisfies $W^{\dagger}=\gamma^{0}W\gamma^{0}$
and therefore it can be expanded in terms of the generators of Clifford
algebra $\Gamma_{i}=\{1_{4},i\gamma^{5},\gamma^{\mu},\gamma^{5}\gamma^{\mu},(1/2)\sigma^{\mu\nu}\}$
as 
\begin{eqnarray}
W(x,p) & = & \frac{1}{4}\left(1_{4}\mathcal{F}+i\gamma^{5}\mathcal{P}+\gamma^{\mu}\mathcal{V}_{\mu}+\gamma^{5}\gamma^{\mu}\mathcal{A}_{\mu}+\frac{1}{2}\sigma^{\mu\nu}\mathcal{S}_{\mu\nu}\right),\label{eq:decomposition}
\end{eqnarray}
where $\sigma^{\mu\nu}=(i/2)[\gamma^{\mu},\gamma^{\nu}]$, and all
expansion coefficients $\{\mathcal{F},\mathcal{P},\mathcal{V}^{\mu},\mathcal{A}^{\nu},\mathcal{S}^{\mu\nu}\}$
are real functions in phase space. Each function contains an equilibrium
and an off-equilibrium part, for example, 
\begin{equation}
\mathcal{F}=\mathcal{F}_{\text{eq}}+\delta\mathcal{F}.
\end{equation}
So $W_{\text{eq}}(x,p)$ also gives its equilibrium components $\{\mathcal{F}_{\text{eq}},\mathcal{P}_{\text{eq}},\mathcal{V}_{\text{eq}}^{\mu},\mathcal{A}_{\text{eq}}^{\nu},\mathcal{S}_{\text{eq}}^{\mu\nu}\}$
similarly as in Eq. (\ref{eq:decomposition}).


Projecting Eq. (\ref{eq:equilibrium equation}) onto $\Gamma_{i}$
and decoupling real and imaginary parts, we obtain the following set
of partial differential equations 
\begin{eqnarray}
p_{\mu}\mathcal{V}_{\text{eq}}^{\mu}-m\mathcal{F}_{\text{eq}} & = & 0,\label{eq:Feq}\\
\frac{\hbar}{2}\left(\partial_{x}^{\mu}-F_{0}^{\mu\nu}\partial_{p\nu}\right)\mathcal{A}_{\mu}^{\text{eq}}+m\mathcal{P}_{\text{eq}} & = & 0,\label{eq:Peq}\\
p_{\mu}\mathcal{F}_{\text{eq}}-\frac{\hbar}{2}\left(\partial_{x}^{\nu}-F_{0}^{\nu\alpha}\partial_{p\alpha}\right)\mathcal{S}_{\nu\mu}^{\text{eq}}-m\mathcal{V}_{\mu}^{\text{eq}} & = & 0,\label{eq:Vmueq}\\
-\frac{\hbar}{2}\left(\partial_{x\mu}-F_{0,\mu\nu}\partial_{p}^{\nu}\right)\mathcal{P}_{\text{eq}}+\frac{1}{2}\epsilon_{\mu\nu\alpha\beta}p^{\nu}\mathcal{S}_{\text{eq}}^{\alpha\beta}+m\mathcal{A}_{\mu}^{\text{eq}} & = & 0,\label{eq:Amueq}\\
\frac{\hbar}{2}\left(\partial_{x\mu}-F_{0,\mu\alpha}\partial_{p}^{\alpha}\right)\mathcal{V}_{\nu}^{\text{eq}}-\frac{\hbar}{2}\left(\partial_{x\nu}-F_{0,\nu\alpha}\partial_{p}^{\alpha}\right)\mathcal{V}_{\mu}^{\text{eq}}\nonumber \\
-\epsilon_{\mu\nu\alpha\beta}p^{\alpha}\mathcal{A}_{\text{eq}}^{\beta}-m\mathcal{S}_{\mu\nu}^{\text{eq}} & = & 0,\label{eq:Smunueq}\\
\hbar\left(\partial_{x}^{\mu}-F_{0}^{\mu\nu}\partial_{p\nu}\right)\mathcal{V}_{\mu}^{\text{eq}} & = & 0,\label{eq:Cons Vmueq}\\
p^{\mu}\mathcal{A}_{\mu}^{\text{eq}} & = & 0,\label{eq:Cons Amueq}\\
\frac{\hbar}{2}\left(\partial_{x\mu}-F_{0,\mu\nu}\partial_{p}^{\nu}\right)\mathcal{F}_{\text{eq}}+p^{\nu}\mathcal{S}_{\nu\mu}^{\text{eq}} & = & 0,\label{eq:Cons Smunueq}\\
p_{\mu}\mathcal{P}_{\text{eq}}+\frac{\hbar}{4}\epsilon_{\mu\nu\alpha\beta}\left(\partial_{x}^{\nu}-F_{0}^{\nu\rho}\partial_{p\rho}\right)\mathcal{S}_{\text{eq}}^{\alpha\beta} & = & 0,\label{eq:Cons Peq}\\
p_{[\mu}\mathcal{V}_{\nu]}^{\text{eq}}+\frac{\hbar}{2}\epsilon_{\mu\nu\alpha\beta}\left(\partial_{x}^{\alpha}-F_{0}^{\alpha\rho}\partial_{p\rho}\right)\mathcal{A}_{\text{eq}}^{\beta} & = & 0.\label{eq:Cons VAeq}
\end{eqnarray}
With the help of Eqs. (\ref{eq:Feq}), (\ref{eq:Peq}), and (\ref{eq:Smunueq}),
we are able to express $\mathcal{F}_{\text{eq}}$, $\mathcal{P}_{\text{eq}}$,
and $\mathcal{S}_{\text{eq}}^{\mu\nu}$ in terms of the remaining
components $\mathcal{V}_{\text{eq}}^{\mu}$ and $\mathcal{A}_{\text{eq}}^{\mu}$
as 
\begin{eqnarray}
\mathcal{F}_{\text{eq}} & = & \frac{1}{m}p_{\mu}\mathcal{V}_{\text{eq}}^{\mu},\nonumber \\
\mathcal{P}_{\text{eq}} & = & -\frac{\hbar}{2m}\left(\partial_{x}^{\mu}-F_{0}^{\mu\nu}\partial_{p\nu}\right)\mathcal{A}_{\mu}^{\text{eq}},\nonumber \\
\mathcal{S}_{\mu\nu}^{\text{eq}} & = & \frac{\hbar}{2m}\left(\partial_{x\mu}-F_{0,\mu\alpha}\partial_{p}^{\alpha}\right)\mathcal{V}_{\nu}^{\text{eq}}-\frac{\hbar}{2m}\left(\partial_{x\nu}-F_{0,\nu\alpha}\partial_{p}^{\alpha}\right)\mathcal{V}_{\mu}^{\text{eq}}\nonumber \\
 &  & -\frac{1}{m}\epsilon_{\mu\nu\alpha\beta}p^{\alpha}\mathcal{A}_{\text{eq}}^{\beta}.
\end{eqnarray}
Substituting above relations into Eqs. (\ref{eq:Cons Smunueq}) and
(\ref{eq:Cons Peq}), we obtain the kinetic equations for $\mathcal{V}_{\text{eq}}^{\mu}$
and $\mathcal{A}_{\text{eq}}^{\mu}$, 
\begin{eqnarray}
p^{\nu}\left(\partial_{x\nu}-F_{0,\nu\alpha}\partial_{p}^{\alpha}\right)\mathcal{V}_{\mu}^{\text{eq}}-F_{0,\mu\nu}\mathcal{V}_{\text{eq}}^{\nu} & = & 0,\nonumber \\
p^{\nu}\left(\partial_{x\nu}-F_{0,\nu\alpha}\partial_{p}^{\alpha}\right)\mathcal{A}_{\mu}^{\text{eq}}-F_{0,\mu\nu}\mathcal{A}_{\text{eq}}^{\nu} & = & 0.\label{eq:Boltzmann-equations}
\end{eqnarray}
On the other hand, substituting $\mathcal{F}_{\text{eq}}$, $\mathcal{P}_{\text{eq}}$,
and $\mathcal{S}_{\text{eq}}^{\mu\nu}$ into Eqs. (\ref{eq:Vmueq})
and (\ref{eq:Amueq}), we obtain the mass-shell constraints 
\begin{eqnarray}
0 & = & \left(p^{2}-m^{2}\right)\mathcal{V}_{\mu}^{\text{eq}}+\frac{\hbar}{2m}\epsilon_{\mu\nu\alpha\beta}F_{0}^{\nu\alpha}\mathcal{A}_{\text{eq}}^{\beta},\nonumber \\
0 & = & \left(p^{2}-m^{2}\right)\mathcal{A}_{\mu}^{\text{eq}}+\frac{\hbar}{2m}\epsilon_{\mu\nu\alpha\beta}F_{0}^{\nu\alpha}\mathcal{V}_{\text{eq}}^{\beta},\label{eq:mass-shell-constraints}
\end{eqnarray}
which indicate that the particle's mass-shell is modified by the magnetic
field.


The Wigner function in a constant magnetic field has been given in
Ref. \citep{Sheng:2017lfu,Sheng:2019ujr} in a non-covariant form.
It is straightforward to generalize the results to the covariant form,
\begin{eqnarray}
\mathcal{V}_{\text{eq}}^{\mu} & = & \sum_{n=0}^{\infty}\left[p^{\mu}-\left(1-\frac{2n\hbar B_{0}}{p_{T}^{2}}\right)\Xi^{\mu\nu}p_{\nu}\right]V_{n}(p)\Lambda_{+}^{(n)}(p_{T}),\nonumber \\
\mathcal{A}_{\text{eq}}^{\mu} & = & \left[(u\cdot p)b^{\mu}-(b\cdot p)u^{\mu}\right]\sum_{n=0}^{\infty}V_{n}(p)\Lambda_{-}^{(n)}(p_{T}),\label{eq:equilibirum-solutions}
\end{eqnarray}
where $b^{\mu}$ is the direction of the magnetic field, $\Xi^{\mu\nu}$
is the projection operator in Eq. (\ref{eq:projection operator}),
$p_{T}=\sqrt{-p_{\alpha}\Xi^{\alpha\beta}p_{\beta}}$ is the magnitude
of the transverse momentum, and $\Lambda_{\pm}^{(n)}(p_{T})$ are
defined as 
\begin{equation}
\Lambda_{\pm}^{(n)}(p_{T})=(-1)^{n}\left[L_{n}\left(\frac{2p_{T}^{2}}{\hbar B_{0}}\right)\mp L_{n-1}\left(\frac{2p_{T}^{2}}{\hbar B_{0}}\right)\right]\exp\left(-\frac{p_{T}^{2}}{\hbar B_{0}}\right),\label{eq:Lambda_pm}
\end{equation}
where $L_{n}(x)$ with $n\geq0$ are Laguerre polynomials and $L_{-1}(x)=0$.
In Eq. (\ref{eq:equilibirum-solutions}), $V_{n}(p)$ is a function
that only depends on $u\cdot p$ and $b\cdot p$, 
\begin{eqnarray}
V_{n}(p) & \equiv & \frac{2}{(2\pi)^{3}}\left(2-\delta_{n0}\right)\delta\left[(u\cdot p)^{2}-(b\cdot p)^{2}-m^{2}-2n\hbar B_{0}\right]\nonumber \\
 &  & \times\left\{ f_{\text{eq}}^{(+)}(u\cdot p)\theta(u\cdot p)+f_{\text{eq}}^{(-)}(-u\cdot p)\theta(-u\cdot p)\right\} .\label{eq:Vn}
\end{eqnarray}
where $f_{\text{eq}}^{(\pm)}$ are Fermi-Dirac distributions at local
temperature $T=1/\beta$ and chemical potential $\mu=\alpha T$,
\begin{equation}
f_{\text{eq}}^{(\pm)}(u\cdot p)=\frac{1}{1+\exp\left[\beta(u\cdot p)\mp\alpha\right]}.\label{eq:equilibrium-distribution}
\end{equation}
The space-time dependence of $\mathcal{V}_{\text{eq}}^{\mu}$ and
$\mathcal{A}_{\text{eq}}^{\mu}$ is through $\beta(x)$, $u^{\mu}(x)$,
and $\mu(x)$.


It can be easily checked that $\mathcal{V}_{\text{eq}}^{\mu}$ and
$\mathcal{A}_{\text{eq}}^{\mu}$ in Eq. (\ref{eq:equilibirum-solutions})
fulfill the mass-shell conditions in (\ref{eq:mass-shell-constraints}),
and $\mathcal{V}_{\text{eq}}^{\mu}$ and $\mathcal{A}_{\text{eq}}^{\mu}$
should also satisfy Eqs. (\ref{eq:Cons Vmueq}), (\ref{eq:Cons Amueq}),
(\ref{eq:Cons VAeq}), and (\ref{eq:Boltzmann-equations}), leading
to following constraints 
\begin{eqnarray}
\partial_{x\mu}\left\{ \left[p^{\mu}-\left(1-\frac{2n\hbar B_{0}}{p_{T}^{2}}\right)\Xi^{\mu\nu}p_{\nu}\right]V_{n}(p)\Lambda_{+}^{(n)}(p_{T})\right\}  & = & 0,\nonumber \\
\epsilon_{\mu\nu\alpha\beta}\partial_{x}^{\alpha}\left\{ \left[(u\cdot p)b^{\beta}-(b\cdot p)u^{\beta}\right]V_{n}(p)\Lambda_{-}^{(n)}(p_{T})\right\}  & = & 0,\nonumber \\
p^{\nu}\partial_{x\nu}\left\{ \left[p^{\mu}-\left(1-\frac{2n\hbar B_{0}}{p_{T}^{2}}\right)\Xi^{\mu\rho}p_{\rho}\right]V_{n}(p)\Lambda_{+}^{(n)}(p_{T})\right\}  & = & 0.\label{eq:Boltzmann equations-1}
\end{eqnarray}
Since $\Lambda_{\pm}^{(n)}(p_{T})$ are independent of space-time
coordinates, the constraints in Eq. (\ref{eq:Boltzmann equations-1})
can be cast into those for equilibrium distributions 
\begin{eqnarray}
\left[p^{\mu}-\left(1-\frac{2n\hbar B_{0}}{p_{T}^{2}}\right)\Xi^{\mu\nu}p_{\nu}\right]\partial_{x\mu}f_{\text{eq}}^{(\pm)}(x;u\cdot p)\nonumber \\
-\left(1-\frac{2n\hbar B_{0}}{p_{T}^{2}}\right)\left(\partial_{x\mu}\Xi^{\mu\nu}p_{\nu}\right)f_{\text{eq}}^{(\pm)}(x;u\cdot p) & = & 0,
\end{eqnarray}
\begin{equation}
\epsilon_{\mu\nu\alpha\beta}\partial_{x}^{\alpha}\left\{ \left[(u\cdot p)b^{\beta}-(b\cdot p)u^{\beta}\right]f_{\text{eq}}^{(\pm)}(x;u\cdot p)\right\} =0,
\end{equation}
\begin{eqnarray}
\left[p^{\mu}-\left(1-\frac{2n\hbar B_{0}}{p_{T}^{2}}\right)\Xi^{\mu\rho}p_{\rho}\right]p^{\nu}\partial_{x\nu}f_{\text{eq}}^{(\pm)}(x;u\cdot p)\nonumber \\
-\left(1-\frac{2n\hbar B_{0}}{p_{T}^{2}}\right)p^{\nu}\left(\partial_{x\nu}\Xi^{\mu\rho}p_{\rho}\right)f_{\text{eq}}^{(\pm)}(x;u\cdot p) & = & 0.
\end{eqnarray}
Substituting $f_{\text{eq}}^{(\pm)}$ in Eq. (\ref{eq:equilibrium-distribution})
into these equations, we find that $\alpha$ should be a space-time
constant, and $\beta u_{\mu}$ should satisfy the Killing condition,
namely, 
\begin{eqnarray}
\partial_{x}^{\mu}\alpha & = & 0,\nonumber \\
\partial_{\mu}^{x}(\beta u_{\nu})+\partial_{\nu}^{x}(\beta u_{\mu}) & = & 0.\label{eq:global-eq-cond}
\end{eqnarray}
These are also conditions for global equilibrium without EM fields.
In addition to the conditions in (\ref{eq:global-eq-cond}), we also
obtain 
\begin{eqnarray}
 &  & \partial_{b}u^{\mu}=0,\;\;\;\;\partial_{\perp}^{\nu}u^{\mu}=0,\nonumber \\
 &  & \partial_{b}b^{\mu}=0,\;\;\;\;\partial_{\perp}^{\nu}b^{\mu}=0,\nonumber \\
 &  & \partial_{\perp}^{\mu}\beta=0,\;\;\;\;\Xi_{\nu}^{\mu}\partial_{t}b^{\nu}=0,
\end{eqnarray}
where $\partial_{t}\equiv u_{\mu}\partial_{x}^{\mu}$, $\partial_{b}\equiv-b_{\mu}\partial_{x}^{\mu}$,
and $\partial_{\perp}^{\mu}\equiv\Xi_{\nu}^{\mu}\partial_{x}^{\nu}$.
These constraints are satisfied by that $\beta$, $b^{\mu}$, and
$u^{\mu}$ are all space-time constants.


\section{General equations of Wigner functions: off-equilibrium \label{sec:off-equilibrium Wigner function}}

The off-equilibrium part of the Wigner function $\delta W$ satisfies
Eq. (\ref{eq:equation_for_deviation}). Similar to Eq. (\ref{eq:decomposition}),
we can decompose $\delta W$ in terms of $\Gamma_{i}$, then from
Eq. (\ref{eq:equation_for_deviation}) we derive the equations for
Clifford components,
\begin{eqnarray}
p_{\mu}\delta\mathcal{V}^{\mu}-m\delta\mathcal{F} & = & 0,\label{eq:deltaF}\\
\frac{\hbar}{2}\left(\partial_{x}^{\mu}-F_{0}^{\mu\nu}\partial_{p\nu}+\frac{1}{\tau}u^{\mu}\right)\delta\mathcal{A}_{\mu}+m\delta\mathcal{P} & = & \frac{\hbar}{2}f^{\mu\nu}\partial_{p\nu}\mathcal{A}_{\mu}^{\text{eq}},\label{eq:deltaP}\\
p_{\mu}\delta\mathcal{F}-\frac{\hbar}{2}\left(\partial_{x}^{\nu}-F_{0}^{\nu\rho}\partial_{p\rho}+\frac{1}{\tau}u^{\nu}\right)\delta\mathcal{S}_{\nu\mu}-m\delta\mathcal{V}_{\mu} & = & \frac{\hbar}{2}f^{\nu\rho}\partial_{p\rho}\mathcal{S}_{\mu\nu}^{\text{eq}},\label{eq:delta Vmu}\\
-\frac{\hbar}{2}\left(\partial_{x\mu}-F_{0,\mu\nu}\partial_{p}^{\nu}+\frac{1}{\tau}u_{\mu}\right)\delta\mathcal{P}+\frac{1}{2}\epsilon_{\mu\nu\alpha\beta}p^{\nu}\delta\mathcal{S}^{\alpha\beta}+m\delta\mathcal{A}_{\mu} & = & -\frac{\hbar}{2}f_{\mu\nu}\partial_{p}^{\nu}\mathcal{P}^{eq},\label{eq:delta Amu}\\
\frac{\hbar}{2}\left(\partial_{x\mu}-F_{0,\mu\rho}\partial_{p}^{\rho}+\frac{1}{\tau}u_{\mu}\right)\delta\mathcal{V}_{\nu}\label{eq:delta Smunu}\\
-\frac{\hbar}{2}\left(\partial_{x\nu}-F_{0,\nu\rho}\partial_{p}^{\rho}+\frac{1}{\tau}u_{\nu}\right)\delta\mathcal{V}_{\mu}-\epsilon_{\mu\nu\alpha\beta}p^{\alpha}\delta\mathcal{A}^{\beta}-m\delta\mathcal{S}_{\mu\nu} & = & \frac{\hbar}{2}f_{\mu\rho}\partial_{p}^{\rho}\mathcal{V}_{\nu}^{\text{eq}}-\frac{\hbar}{2}f_{\nu\rho}\partial_{p}^{\rho}\mathcal{V}_{\mu}^{\text{eq}},\nonumber \\
\hbar\left(\partial_{x}^{\mu}-F_{0}^{\mu\nu}\partial_{p\nu}+\frac{1}{\tau}u^{\mu}\right)\delta\mathcal{V}_{\mu} & = & \hbar f^{\mu\nu}\partial_{p\nu}\mathcal{V}_{\mu}^{\text{eq}},\label{eq:Cons Vmu}\\
p^{\mu}\delta\mathcal{A}_{\mu} & = & 0,\label{eq:Cons Amu}\\
\frac{\hbar}{2}\left(\partial_{x\mu}-F_{0,\mu\nu}\partial_{p}^{\nu}+\frac{1}{\tau}u_{\mu}\right)\delta\mathcal{F}+p^{\nu}\delta\mathcal{S}_{\nu\mu} & = & \frac{\hbar}{2}f_{\mu\rho}\partial_{p}^{\rho}\mathcal{F}_{\text{eq}},\label{eq:Cons Smunu}\\
p_{\mu}\delta\mathcal{P}+\frac{\hbar}{4}\epsilon_{\mu\nu\alpha\beta}\left(\partial_{x}^{\nu}-F_{0}^{\nu\rho}\partial_{p\rho}+\frac{1}{\tau}u^{\nu}\right)\delta\mathcal{S}^{\alpha\beta} & = & \frac{\hbar}{4}\epsilon_{\mu\nu\alpha\beta}f^{\nu\rho}\partial_{p\text{\ensuremath{\rho}}}\mathcal{S}_{\text{eq}}^{\alpha\beta},\label{eq:Cons P}\\
p_{\mu}\delta\mathcal{V}_{\nu}-p_{\nu}\delta\mathcal{V}_{\mu}+\frac{\hbar}{2}\epsilon_{\mu\nu\alpha\beta}\left(\partial_{x}^{\alpha}-F_{0}^{\alpha\rho}\partial_{p\rho}+\frac{1}{\tau}u^{\alpha}\right)\delta\mathcal{A}^{\beta} & = & \frac{\hbar}{2}\epsilon_{\mu\nu\alpha\beta}f^{\alpha\rho}\partial_{p\text{\ensuremath{\rho}}}\mathcal{A}_{\text{eq}}^{\beta}.\label{eq:Cons VA}
\end{eqnarray}
Taking the vector and axial-vector components $\delta\mathcal{V}_{\mu}$
and $\delta\mathcal{A}_{\mu}$ as basic ones, the remaining components
are then given by
\begin{eqnarray}
\delta\mathcal{F} & = & \frac{1}{m}p_{\mu}\delta\mathcal{V}^{\mu},\label{eq:F}\\
\delta\mathcal{P} & = & \frac{\hbar}{2m}f^{\mu\nu}\partial_{p\nu}\mathcal{A}_{\mu}^{\text{eq}}-\frac{\hbar}{2m}\left(\partial_{x}^{\mu}-F_{0}^{\mu\nu}\partial_{p\nu}+\frac{1}{\tau}u^{\mu}\right)\delta\mathcal{A}_{\mu},\label{eq:P}\\
\delta\mathcal{S}_{\mu\nu} & = & \frac{\hbar}{2m}\left(\partial_{x\mu}-F_{0,\mu\rho}\partial_{p}^{\rho}+\frac{1}{\tau}u_{\mu}\right)\delta\mathcal{V}_{\nu}-\frac{\hbar}{2m}\left(\partial_{x\nu}-F_{0,\nu\rho}\partial_{p}^{\rho}+\frac{1}{\tau}u_{\nu}\right)\delta\mathcal{V}_{\mu}\nonumber \\
 &  & -\frac{1}{m}\epsilon_{\mu\nu\alpha\beta}p^{\alpha}\delta\mathcal{A}^{\beta}-\frac{\hbar}{2m}f_{\mu\rho}\partial_{p}^{\rho}\mathcal{V}_{\nu}^{\text{eq}}+\frac{\hbar}{2m}f_{\nu\rho}\partial_{p}^{\rho}\mathcal{V}_{\mu}^{\text{eq}}.\label{eq:Smunu}
\end{eqnarray}
Substituting $\delta\mathcal{F}$, $\delta\mathcal{P}$, and $\delta\mathcal{S}_{\mu\nu}$
into Eqs. (\ref{eq:delta Vmu}) and (\ref{eq:delta Amu}), we obtain
the mass-shell constraints for $\delta\mathcal{V}_{\mu}$ and $\delta\mathcal{A}_{\mu}$,
\begin{eqnarray}
(p^{2}-m^{2})\delta\mathcal{V}_{\mu}+\frac{\hbar}{2}\epsilon_{\mu\nu\alpha\beta}F_{0}^{\nu\alpha}\delta\mathcal{A}^{\beta} & = & -\frac{\hbar}{2}\epsilon_{\mu\nu\alpha\beta}f^{\nu\alpha}\mathcal{A}_{\text{eq}}^{\beta}+\mathcal{O}(\hbar^{2}),\label{eq:mass-shell-1}\\
(p^{2}-m^{2})\delta\mathcal{A}_{\mu}-\frac{\hbar}{2}\epsilon_{\mu\nu\alpha\beta}F_{0}^{\alpha\nu}\delta\mathcal{V}^{\beta} & = & -\frac{\hbar}{2}\epsilon_{\mu\nu\alpha\beta}f^{\nu\alpha}\mathcal{V}_{\text{eq}}^{\beta}+\mathcal{O}(\hbar^{2}).\label{eq:mass-shell-2}
\end{eqnarray}
On the other hand, the kinetic or Boltzmann equations for $\delta\mathcal{V}_{\mu}$
and $\delta\mathcal{A}_{\mu}$ are derived from Eqs. (\ref{eq:Cons Smunu})
and (\ref{eq:Cons VA}) 
\begin{eqnarray}
 &  & \hbar p^{\nu}\left(\partial_{x\nu}-F_{0,\nu\rho}\partial_{p}^{\rho}+\frac{1}{\tau}u_{\nu}\right)\delta\mathcal{V}_{\mu}-\hbar F_{0,\mu\nu}\delta\mathcal{V}^{\nu}\nonumber \\
 & = & \hbar f_{\mu\nu}\mathcal{V}_{\text{eq}}^{\nu}+\hbar p^{\nu}f_{\nu\rho}\partial_{p}^{\rho}\mathcal{V}_{\mu}^{\text{eq}}+\mathcal{O}(\hbar^{2}),\label{eq:Boltzmann-1}\\
 &  & \hbar p_{\nu}\left(\partial_{x}^{\nu}-F_{0}^{\nu\rho}\partial_{p\rho}+\frac{1}{\tau}u^{\nu}\right)\delta\mathcal{A}_{\mu}-\hbar F_{0,\mu\nu}\delta\mathcal{A}^{\nu}\nonumber \\
 & = & \hbar f_{\mu\nu}\mathcal{A}_{\text{eq}}^{\nu}+\hbar p^{\nu}f_{\nu\rho}\partial_{p}^{\rho}\mathcal{A}_{\mu}^{\text{eq}}+\mathcal{O}(\hbar^{2}),\label{eq:Boltzmann-2}
\end{eqnarray}
where we have dropped terms in $\mathcal{O}(\hbar^{2})$ and higher
orders. Since we considered the case of the strong background magnetic
field $B_{0}$, we treat $\hbar B_{0}$ as the zeroth order quantity
in power counting of $\hbar$. As the result, the equilibrium solutions
in Eq. (\ref{eq:equilibirum-solutions}) are the zeroth order contributions.
Consequently, $\delta\mathcal{V}_{\mu}$ and $\delta\mathcal{A}_{\mu}$
obtained by solving Eq. (\ref{eq:Boltzmann-1}) and (\ref{eq:Boltzmann-2})
are also the zeroth order contributions. In the following discussions,
we will set $\hbar=1$ for simplicity.


In the strong magnetic field, particle's energy spectra are described
by the Landau levels. However, the perturbation $f^{\mu\nu}$ will
modify the energy spectra, as shown in Eqs. (\ref{eq:mass-shell-1})
and (\ref{eq:mass-shell-2}). Since the perturbation $f^{\mu\nu}$
is assumed to be much weaker than $F_{0}^{\mu\nu}$, one can safely
expect that the deviation from Landau levels is negligible. Therefore
in the following discussion, we neglect the mass-shell constraints
for $\delta\mathcal{V}_{\mu}$ and $\delta\mathcal{A}_{\mu}$ in Eqs.
(\ref{eq:mass-shell-1}) and (\ref{eq:mass-shell-2}), and focusing
on the Boltzmann equations (\ref{eq:Boltzmann-1}) and (\ref{eq:Boltzmann-2}).
For simplicity, we set $B_{0}$, $\beta$, $b^{\mu}$, and $u^{\mu}$
as spacetime-independent quantities, and assume that $f^{\mu\nu}$
only depend on the proper time $t\equiv u\cdot x$. We decompose the
perturbation $f^{\mu\nu}$ into the electric and magnetic part as
\begin{equation}
f^{\mu\nu}=\delta E^{\mu}u^{\nu}-\delta E^{\nu}u^{\mu}+\epsilon^{\mu\nu\alpha\beta}u_{\alpha}\delta B_{\beta}.\label{eq:decomposing_fmunu}
\end{equation}
In Eqs. (\ref{eq:Boltzmann-1}) and (\ref{eq:Boltzmann-2}), $f^{\mu\nu}$
plays the role of a source term for $\delta\mathcal{V}_{\mu}$ and
$\delta\mathcal{A}_{\mu}$. In general, the obtained $\delta\mathcal{V}_{\mu}$
and $\delta\mathcal{A}_{\mu}$ can be expressed as linear responses
to $\delta E^{\mu}$ and $\delta B^{\mu}$. The induced vector and
axial vector currents, defined as momentum integrals of $\delta\mathcal{V}^{\mu}$
and $\delta\mathcal{A}^{\mu}$, are given by 
\begin{eqnarray}
\delta J^{\mu} & \equiv & \int d^{4}p\,\delta\mathcal{V}^{\mu}=\sigma_{E}^{\mu\nu}\delta E_{\nu}+\sigma_{B}^{\mu\nu}\delta B_{\nu},\nonumber \\
\delta J_{A}^{\mu} & \equiv & \int d^{4}p\,\delta\mathcal{A}^{\mu}=\chi_{A,E}^{\mu\nu}\delta E_{\nu}+\chi_{A,B}^{\mu\nu}\delta B_{\nu},
\end{eqnarray}
where the linear responses are described by conductivities $\sigma_{E}^{\mu\nu}$,
$\sigma_{B}^{\mu\nu}$, $\chi_{A,E}^{\mu\nu}$, and $\chi_{A,B}^{\mu\nu}$.
Since we assumed that $\delta E^{\mu}$ and $\delta B^{\mu}$ are
functions of $t$, these conductivitites are also proper time-dependent
quantities. In the following two sections, we will show 
\begin{eqnarray}
\sigma_{E}^{\mu\nu} & = & \Xi^{\mu\nu}\sigma_{\perp}-b^{\mu}b^{\nu}\sigma_{\parallel}+\epsilon^{\mu\nu\rho\sigma}u_{\rho}b_{\sigma}\sigma_{H},\nonumber \\
\sigma_{B}^{\mu\nu} & = & 0,\nonumber \\
\chi_{A,E}^{\mu\nu} & = & -u^{\mu}b^{\nu}\chi_{A},\nonumber \\
\chi_{A,B}^{\mu\nu} & = & \Xi^{\mu\nu}\chi_{\parallel}+\epsilon^{\mu\nu\rho\sigma}u_{\rho}b_{\sigma}\chi_{\perp},
\end{eqnarray}
where $\sigma_{\perp}$, $\sigma_{\parallel}$ are Ohm conductivities
in the transverse and longitudinal directions with respect to the
direction of the background magnetic field, $\sigma_{H}$ is the Hall
conductivity, $\chi_{A}$ is the coefficient for the axial-charge
induced by parallel electric and background magnetic field, and $\chi_{\parallel}$
and $\chi_{\perp}$ describe the axial current induced by the perturbative
magnetic field $\delta B^{\mu}$. The explicit expressions for these
conductivities will be given in following sections.


\section{Vector current induced by perturbation fields \label{sec:Vector-current-induced}}

\subsection{Analytical solutions}

In this section, we solve the Boltzmann equation (\ref{eq:Boltzmann-1})
to obtain the vector current. With $F_{0}^{\mu\nu}$ in Eq. (\ref{eq:background field}),
the Boltzmann equation (\ref{eq:Boltzmann-1}) can be written as 
\begin{equation}
\left[(u\cdot p)\left(\frac{d}{dt}+\frac{1}{\tau}\right)-B_{0}\epsilon^{\nu\lambda\rho\sigma}p_{\nu}^{\perp}\partial_{\lambda}^{p,\perp}u_{\rho}b_{\sigma}\right]\delta\mathcal{V}_{\mu}-B_{0}\epsilon_{\mu\nu\rho\sigma}u^{\rho}b^{\sigma}\delta\mathcal{V}^{\nu}=f_{\mu\nu}\mathcal{V}_{\text{eq}}^{\nu}+p^{\nu}f_{\nu\rho}\partial_{p}^{\rho}\mathcal{V}_{\mu}^{\text{eq}}.\label{eq:Boltzmann equation deltaVmu}
\end{equation}
where we made approximation that $\delta\mathcal{V}_{\mu}$ only depends
on the proper time $t\equiv u\cdot x$ because $f^{\mu\nu}$ is assumed
to be only $t$-dependent. The transverse momentum and the transverse
momentum derivative are defined with the help of the projection operator
$\Xi^{\mu\nu}$ in Eq. (\ref{eq:projection operator}),
\begin{equation}
p_{\perp}^{\mu}\equiv\Xi^{\mu\nu}p_{\nu},\ \ \partial_{p,\perp}^{\mu}\equiv\Xi^{\mu\nu}\partial_{\nu}^{p},
\end{equation}
while the longitudinal terms are 
\begin{equation}
p_{\parallel}^{\mu}\equiv p^{\mu}-p_{\perp}^{\mu},\ \ \partial_{\parallel,p}^{\mu}\equiv\partial_{p}^{\mu}-\partial_{\perp,p}^{\mu}.
\end{equation}
The vectors $u^{\mu}$, $b^{\mu}$, $p_{\perp}^{\mu}$, and $\epsilon^{\mu\nu\rho\sigma}p_{\nu}^{\perp}u_{\rho}b_{\sigma}$
are orthogonal to each other when $p_{\perp}^{\mu}\neq0$ and therefore
they form a complete set of bases $\{e_{i}^{\mu},i=1,2,3,4\}\equiv\{u^{\mu},b^{\mu},p_{\perp}^{\mu},\epsilon^{\mu\nu\rho\sigma}p_{\nu}^{\perp}u_{\rho}b_{\sigma}\}$
for any four-vector. The vector $\delta\mathcal{V}^{\mu}$ can then
be decomposed as 
\begin{equation}
\delta\mathcal{V}^{\mu}=u^{\mu}\delta\mathcal{V}_{1}-b^{\mu}\delta\mathcal{V}_{2}-\frac{1}{p_{T}^{2}}\left[p_{\perp}^{\mu}\delta\mathcal{V}_{3}^{\perp}+\epsilon^{\mu\nu\rho\sigma}p_{\nu}^{\perp}u_{\rho}b_{\sigma}\delta\mathcal{V}_{4}^{\perp}\right],\label{eq:decomposing-deltaVmu}
\end{equation}
where we have used $p_{T}=\sqrt{-p_{\perp}^{\mu}p_{\mu}^{\perp}}$.
One can extract all components as $\delta\mathcal{V}_{i}=e_{i}^{\mu}\delta\mathcal{V}_{\mu}$.


Substituting $\delta\mathcal{V}_{\mu}$ in Eq. (\ref{eq:decomposing-deltaVmu})
into the Boltzmann equation (\ref{eq:Boltzmann equation deltaVmu}),
we obtain an equation for each component
\begin{equation}
\left[(u\cdot p)\left(\frac{d}{dt}+\frac{1}{\tau}\right)-B_{0}\epsilon^{\nu\lambda\rho\sigma}p_{\nu}^{\perp}\partial_{\lambda}^{p,\perp}u_{\rho}b_{\sigma}\right]\delta\mathcal{V}_{i}=e_{i}^{\mu}\left(f_{\mu\nu}\mathcal{V}_{\text{eq}}^{\nu}+p^{\nu}f_{\nu\rho}\partial_{p}^{\rho}\mathcal{V}_{\mu}^{\text{eq}}\right).\label{eq:equation_delta_Vi}
\end{equation}
The source term on the right-hand-side can be put into the form 
\begin{equation}
e_{i}^{\mu}\left(f_{\mu\nu}\mathcal{V}_{\text{eq}}^{\nu}+p^{\nu}f_{\nu\rho}\partial_{p}^{\rho}\mathcal{V}_{\mu}^{\text{eq}}\right)=C_{i}^{(1)}(t,p_{T},u\cdot p,b\cdot p)+p_{\perp}^{\mu}C_{i,\mu}^{(2)}(t,p_{T},u\cdot p,b\cdot p),\label{eq:source_terms_Ci}
\end{equation}
where $C_{i}^{(1)}$ and $C_{i,\mu}^{(2)}$ are even functions of
$p_{\perp}^{\mu}$. With Eqs. (\ref{eq:equilibirum-solutions}) and
(\ref{eq:source_terms_Ci}), we obtain the source terms that correspond
to $f_{i=1,2}^{(1)}$ and $f_{i=3,4}^{(2)\mu}$ as 
\begin{eqnarray}
C_{1}^{(1)} & = & (u\cdot p)p_{\parallel}^{\nu}f_{\nu\rho}\sum_{n=0}^{\infty}\left[\partial_{\parallel,p}^{\rho}V_{n}(p)\right]\Lambda_{+}^{(n)}(p_{T}),\nonumber \\
C_{2}^{(1)} & = & (b\cdot p)p_{\parallel}^{\nu}f_{\nu\rho}\sum_{n=0}^{\infty}\left[\partial_{\parallel,p}^{\rho}V_{n}(p)\right]\Lambda_{+}^{(n)}(p_{T}),\nonumber \\
C_{3}^{(2)\mu} & = & \sum_{n=0}^{\infty}\Xi^{\mu\nu}f_{\nu\rho}p_{\parallel}^{\rho}\left(1+\frac{2nB_{0}}{p_{T}^{2}}\right)V_{n}(p)\Lambda_{+}^{(n)}(p_{T})\nonumber \\
 &  & -\Xi^{\mu\nu}f_{\nu\rho}\sum_{n=0}^{\infty}2nB_{0}\left[\partial_{\parallel,p}^{\rho}V_{n}(p)\right]\Lambda_{+}^{(n)}(p_{T})\nonumber \\
 &  & -2\Xi^{\mu\nu}f_{\nu\rho}p_{\parallel}^{\rho}\sum_{n=0}^{\infty}2nB_{0}V_{n}(p)\left[\frac{\partial}{\partial p_{T}^{2}}\Lambda_{+}^{(n)}(p_{T})\right],\nonumber \\
C_{4}^{(2)\mu} & = & -\sum_{n=0}^{\infty}\left(1-\frac{2nB_{0}}{p_{T}^{2}}\right)\epsilon^{\mu\nu\rho\sigma}u_{\rho}b_{\sigma}f_{\nu\lambda}p_{\parallel}^{\lambda}V_{n}(p)\Lambda_{+}^{(n)}(p_{T}).\label{eq:source-terms}
\end{eqnarray}


Assuming $\delta\mathcal{V}_{i}(t_{0})=0$ at a given initial moment
$t_{0}$, the solution to Eq. (\ref{eq:equation_delta_Vi}) reads
\begin{equation}
\delta\mathcal{V}_{i}(t,p)=f_{i}^{(1)}(t,p_{T},u\cdot p,b\cdot p)+p_{\perp}^{\mu}f_{i,\mu}^{(2)}(t,p_{T},u\cdot p,b\cdot p),\label{eq:solution_delta_Vi}
\end{equation}
where $f_{i}^{(1)}$ and $f_{i,\mu}^{(2)}$ are obtained from $C_{i}^{(1)}$
and $C_{i}^{(2)\mu}$ as 
\begin{eqnarray}
f_{i}^{(1)}(t) & = & \frac{1}{u\cdot p}\int_{t_{0}}^{t}dt^{\prime}e^{(t^{\prime}-t)/\tau}C_{i}^{(1)}(t^{\prime}),\nonumber \\
f_{i,\mu}^{(2)}(t) & = & -\frac{1}{B_{0}(u\cdot p)}\int_{t_{0}}^{t}dt^{\prime}\sin\left[\frac{B_{0}}{u\cdot p}(t^{\prime}-t)\right]e^{(t^{\prime}-t)/\tau}\nonumber \\
 &  & \times\left[(u\cdot p)\left(\frac{d}{dt^{\prime}}+\frac{1}{\tau}\right)C_{i,\mu}^{(2)}(t^{\prime})+B_{0}\epsilon_{\mu\nu\alpha\beta}u^{\alpha}b^{\beta}C_{i}^{(2)\nu}(t^{\prime})\right].\label{eq:general-solutions}
\end{eqnarray}
Here we have suppressed the dependence of all functions on $p_{T}$,
$u\cdot p$, and $b\cdot p$ for notational simplicity. Substituting
Eq. (\ref{eq:solution_delta_Vi}) into Eq. (\ref{eq:decomposing-deltaVmu}),
we obtain 
\begin{eqnarray}
\delta\mathcal{V}^{\mu} & = & u^{\mu}f_{1}^{(1)}-b^{\mu}f_{2}^{(1)}\nonumber \\
 &  & +\left[u^{\mu}f_{1,\nu}^{(2)}-b^{\mu}f_{2,\nu}^{(2)}\right]p_{\perp}^{\nu}+\frac{1}{p_{T}^{2}}\left[p_{\perp}^{\mu}f_{3}^{(1)}+\epsilon^{\mu\nu\rho\sigma}p_{\nu}^{\perp}u_{\rho}b_{\sigma}f_{4}^{(1)}\right]\nonumber \\
 &  & -\frac{1}{p_{T}^{2}}\left[p_{\perp}^{\mu}p_{\nu}^{\perp}f_{3}^{(2)\nu}+\epsilon^{\mu\nu\rho\sigma}p_{\nu}^{\perp}u_{\rho}b_{\sigma}p_{\lambda}^{\perp}f_{4}^{(2)\lambda}\right].
\end{eqnarray}
Note that $f_{i}^{(1)}$ and $f_{i}^{(2)\mu}$ are even functions
of $p_{\mu}^{\perp}$, thus when calculating the vector current $J^{\mu}(x)$
through integration $\delta\mathcal{V}_{\mu}$ over four-momentum,
only terms with $f_{1}^{(1)}$, $f_{2}^{(1)}$, $f_{3}^{(2)\mu}$
and $f_{4}^{(2)\mu}$ contribute but other terms do not. So we have
\begin{eqnarray}
J^{\mu}(x) & = & u^{\mu}\int d^{4}p\,f_{1}^{(1)}-b^{\mu}\int d^{4}p\,f_{2}^{(1)}\nonumber \\
 &  & +\frac{1}{2}\int d^{4}p\,\left[f_{3}^{(2)\mu}+\epsilon^{\mu\nu\rho\sigma}u_{\rho}b_{\sigma}f_{4,\nu}^{(2)}\right],\label{eq:vector-current}
\end{eqnarray}
where we have used the relation
\begin{equation}
\int d^{4}p\frac{p_{\perp}^{\mu}p_{\nu}^{\perp}}{p_{T}^{2}}f_{i}^{(2)\nu}=-\frac{1}{2}\int d^{4}p\,f_{i}^{(2)\mu}.
\end{equation}
The vector current density $J_{\mu}(x)$ can be calculated using Eqs.
(\ref{eq:general-solutions}), (\ref{eq:vector-current}), and (\ref{eq:source-terms})
and the decomposition of $f^{\mu\nu}$ in Eq. (\ref{eq:decomposing_fmunu}),
\begin{eqnarray}
J^{\mu}(x) & = & -\frac{1}{(2\pi)^{2}}\sum_{n=0}^{\infty}\left(2-\delta_{n0}\right)\int d(b\cdot p)\int_{t_{0}}^{t}dt^{\prime}e^{(t^{\prime}-t)/\tau}\lim_{\varepsilon\rightarrow\varepsilon_{b\cdot p}^{(n)}}\left[f_{\text{eq}}^{(+)}(\varepsilon)+f_{\text{eq}}^{(-)}(\varepsilon)\right]\nonumber \\
 &  & \times\left\{ b^{\mu}b^{\nu}\delta E_{\nu}(t^{\prime})\frac{B_{0}(m^{2}+2nB_{0})}{\varepsilon^{3}}\right.\nonumber \\
 &  & \left.+\left[\Xi^{\mu\nu}\left(\frac{d}{dt^{\prime}}+\frac{1}{\tau}\right)\delta E_{\nu}(t^{\prime})\right]\left(1+\frac{nB_{0}}{\varepsilon}\frac{\partial}{\partial\varepsilon}\right)\sin\left[\frac{B_{0}}{\varepsilon}(t^{\prime}-t)\right]\right\} \nonumber \\
 &  & -\frac{1}{(2\pi)^{2}}\sum_{n=0}^{\infty}\left(2-\delta_{n0}\right)\int d(b\cdot p)\int_{t_{0}}^{t}dt^{\prime}e^{(t^{\prime}-t)/\tau}\lim_{\varepsilon\rightarrow\varepsilon_{b\cdot p}^{(n)}}\left[f_{\text{eq}}^{(+)}(\varepsilon)-f_{\text{eq}}^{(-)}(\varepsilon)\right]\nonumber \\
 &  & \times\frac{B_{0}}{\varepsilon}\epsilon^{\mu\nu\rho\sigma}u_{\rho}b_{\sigma}\delta E_{\nu}(t^{\prime})\left(1+nB_{0}\frac{\partial}{\partial\varepsilon}\frac{1}{\varepsilon}\right)\sin\left[\frac{B_{0}}{\varepsilon}(t^{\prime}-t)\right],
\end{eqnarray}
where $f_{\text{eq}}^{(\pm)}(\varepsilon)\equiv f_{\text{eq}}^{(\pm)}(u\cdot p=\varepsilon)$
are given in Eq. (\ref{eq:equilibrium-distribution}), $\varepsilon_{b\cdot p}^{(n)}\equiv\sqrt{m^{2}+(b\cdot p)^{2}+2nB_{0}}$
is the on-shell energy at $n$-th Landau level, and we have completed
the integrals over $u\cdot p$ and $p_{\perp}^{\mu}$ using explicit
expressions for $\mathcal{V}_{\text{eq}}^{\mu}$, $\Lambda_{\pm}^{(n)}(p_{T})$
and $V_{n}(p)$ in Eqs. (\ref{eq:general-solutions}), (\ref{eq:Lambda_pm}),
and (\ref{eq:Vn}).


Assuming that the electric fields are suddenly switched on at $t_{0}=0$,
$\delta E^{\mu}\propto\theta(t)$, and $\delta E^{\mu}$ is a constant
for $t>0$. Then we can write the vector current as
\begin{equation}
J^{\mu}(x)=\sigma_{\perp}\Xi^{\mu\nu}\delta E_{\nu}-\sigma_{\parallel}b^{\mu}b^{\nu}\delta E_{\nu}+\sigma_{H}\epsilon^{\mu\nu\rho\sigma}u_{\rho}b_{\sigma}\delta E_{\nu},
\end{equation}
where we still used $\delta E_{\nu}$ but to represent the constant
magnitude of the electric field perturbation, $\sigma_{\parallel}$
and $\sigma_{\perp}$ are the electric conductivities in the longitudinal
and transverse direction respectively, and $\sigma_{H}$ is the Hall
conductivity. These conductivities are given by 
\begin{eqnarray}
\sigma_{\parallel} & = & \frac{B_{0}}{(2\pi)^{2}}\mathcal{R}_{1}(t)\sum_{n=0}^{\infty}\left(2-\delta_{n0}\right)\int d(b\cdot p)\nonumber \\
 &  & \times\lim_{\varepsilon\rightarrow\varepsilon_{b\cdot p}^{(n)}}\frac{m^{2}+2nB_{0}}{\varepsilon^{3}}\left[f_{\text{eq}}^{(+)}(\varepsilon)+f_{\text{eq}}^{(-)}(\varepsilon)\right],\label{eq:SigmaPara}\\
\sigma_{\perp} & = & -\frac{1}{(2\pi)^{2}\tau}\sum_{n=0}^{\infty}\left(2-\delta_{n0}\right)\int d(b\cdot p)\nonumber \\
 &  & \times\lim_{\varepsilon\rightarrow\varepsilon_{b\cdot p}^{(n)}}\left[\left(1+\frac{nB_{0}}{\varepsilon}\frac{\partial}{\partial\varepsilon}\right)\mathcal{R}_{2}(\varepsilon,B_{0},t)\right]\left[f_{\text{eq}}^{(+)}(\varepsilon)+f_{\text{eq}}^{(-)}(\varepsilon)\right],\label{eq:SigmaPerp}\\
\sigma_{H} & = & -\frac{B_{0}}{(2\pi)^{2}}\sum_{n=0}^{\infty}\left(2-\delta_{n0}\right)\int d(b\cdot p)\nonumber \\
 &  & \times\lim_{\varepsilon\rightarrow\varepsilon_{b\cdot p}^{(n)}}\frac{1}{\varepsilon}\left[\left(1+nB_{0}\frac{\partial}{\partial\varepsilon}\frac{1}{\varepsilon}\right)\mathcal{R}_{2}(\varepsilon,B_{0},t)\right]\left[f_{\text{eq}}^{(+)}(\varepsilon)-f_{\text{eq}}^{(-)}(\varepsilon)\right].\label{eq:SigmaHall}
\end{eqnarray}
The auxiliary functions $\mathcal{R}_{1}(t)$ and $\mathcal{R}_{2}(\varepsilon,B_{0},t)$
are defined as 
\begin{eqnarray}
\mathcal{R}_{1}(t) & \equiv & \tau(1-e^{-t/\tau}),\nonumber \\
\mathcal{R}_{2}(\varepsilon,B_{0},t) & \equiv & -\frac{\varepsilon\tau}{\varepsilon^{2}+B_{0}^{2}\tau^{2}}\left\{ B_{0}\tau-e^{-t/\tau}\left[B_{0}\tau\cos\left(\frac{B_{0}}{\varepsilon}t\right)+\varepsilon\sin\left(\frac{B_{0}}{\varepsilon}t\right)\right]\right\} ,\label{eq:Auxiliary-functions}
\end{eqnarray}
which controls the time evolution. Since $\ensuremath{f_{\text{eq}}^{(\pm)}\rightarrow0}$
at $\ensuremath{\varepsilon\rightarrow\infty}$, we find that the
sum over Landau levels converges. This ensures that all these conductivities
have finite values.


At the long time limit $t\rightarrow\infty$, the auxiliary functions
$\mathcal{R}_{1}(t)$ and $\mathcal{R}_{2}(\varepsilon,B_{0},t)$
in Eq. (\ref{eq:Auxiliary-functions}) are reduced to $\tau$ and
$-\varepsilon B_{0}\tau^{2}/(\varepsilon^{2}+B_{0}^{2}\tau^{2})$,
respectively. The asymptotic values of conductivitites are 
\begin{eqnarray}
\lim_{t\rightarrow\infty}\sigma_{\parallel} & = & \frac{B_{0}\tau}{(2\pi)^{2}}\sum_{n=0}^{\infty}\left(2-\delta_{n0}\right)\int d(b\cdot p)\lim_{\varepsilon\rightarrow\varepsilon_{b\cdot p}^{(n)}}\frac{m^{2}+2nB_{0}}{\varepsilon^{3}}\left[f_{\text{eq}}^{(+)}(\varepsilon)+f_{\text{eq}}^{(-)}(\varepsilon)\right],\label{eq:SigmaParaStable}\\
\lim_{t\rightarrow\infty}\sigma_{\perp} & = & \frac{B_{0}\tau}{(2\pi)^{2}}\sum_{n=0}^{\infty}\left(2-\delta_{n0}\right)\int d(b\cdot p)\lim_{\varepsilon\rightarrow\varepsilon_{b\cdot p}^{(n)}}\left[\left(1+\frac{nB_{0}}{\varepsilon}\frac{\partial}{\partial\varepsilon}\right)\frac{\varepsilon}{\varepsilon^{2}+B_{0}^{2}\tau^{2}}\right]\nonumber \\
 &  & \times\left[f_{\text{eq}}^{(+)}(\varepsilon)+f_{\text{eq}}^{(-)}(\varepsilon)\right],\label{eq:SigmaPerpStable}\\
\lim_{t\rightarrow\infty}\sigma_{H} & = & \frac{B_{0}^{2}\tau^{2}}{(2\pi)^{2}}\sum_{n=0}^{\infty}\left(2-\delta_{n0}\right)\int d(b\cdot p)\lim_{\varepsilon\rightarrow\varepsilon_{b\cdot p}^{(n)}}\frac{1}{\varepsilon^{2}+B_{0}^{2}\tau^{2}}\left(1-\frac{2nB_{0}}{\varepsilon^{2}+B_{0}^{2}\tau^{2}}\right)\nonumber \\
 &  & \times\left[f_{\text{eq}}^{(+)}(\varepsilon)-f_{\text{eq}}^{(-)}(\varepsilon)\right].\label{eq:SigmaHallStable}
\end{eqnarray}
We observe that $\sigma_{\parallel}(t\rightarrow\infty)$ is always
proportional to the relaxation time $\tau$, which is consistent with
the result in materials \citep{sommerfield1928elektronentheorie}.
For weak magnetic fields, the transverse conductivity $\sigma_{\perp}(t\rightarrow\infty)$
is also proportional to $\tau$, while the Hall conductivity $\sigma_{H}$
is proportional to $\tau^{2}$. Strong magnetic fields break the linear
proportionality between $\sigma_{\perp}(t\rightarrow\infty)$ and
$\tau$, and that between $\sigma_{H}$ and $\tau^{2}$ due to the
term $B_{0}^{2}\tau^{2}$ in the denominators in Eqs. (\ref{eq:SigmaPerpStable})
and (\ref{eq:SigmaHallStable}).


\subsection{Numerical results}

Now we present numerical results for conductivitites in Eqs. (\ref{eq:SigmaPara})-(\ref{eq:SigmaHall}).
In displaying numerical results, we regard the temperature $T$ as
the energy scale so that all quantities and parameters can be made
dimensionless by multiplying powers of $T$. We choose values of dimensionless
parameters in numerical calculation.

\begin{figure}
\includegraphics[scale=0.22]{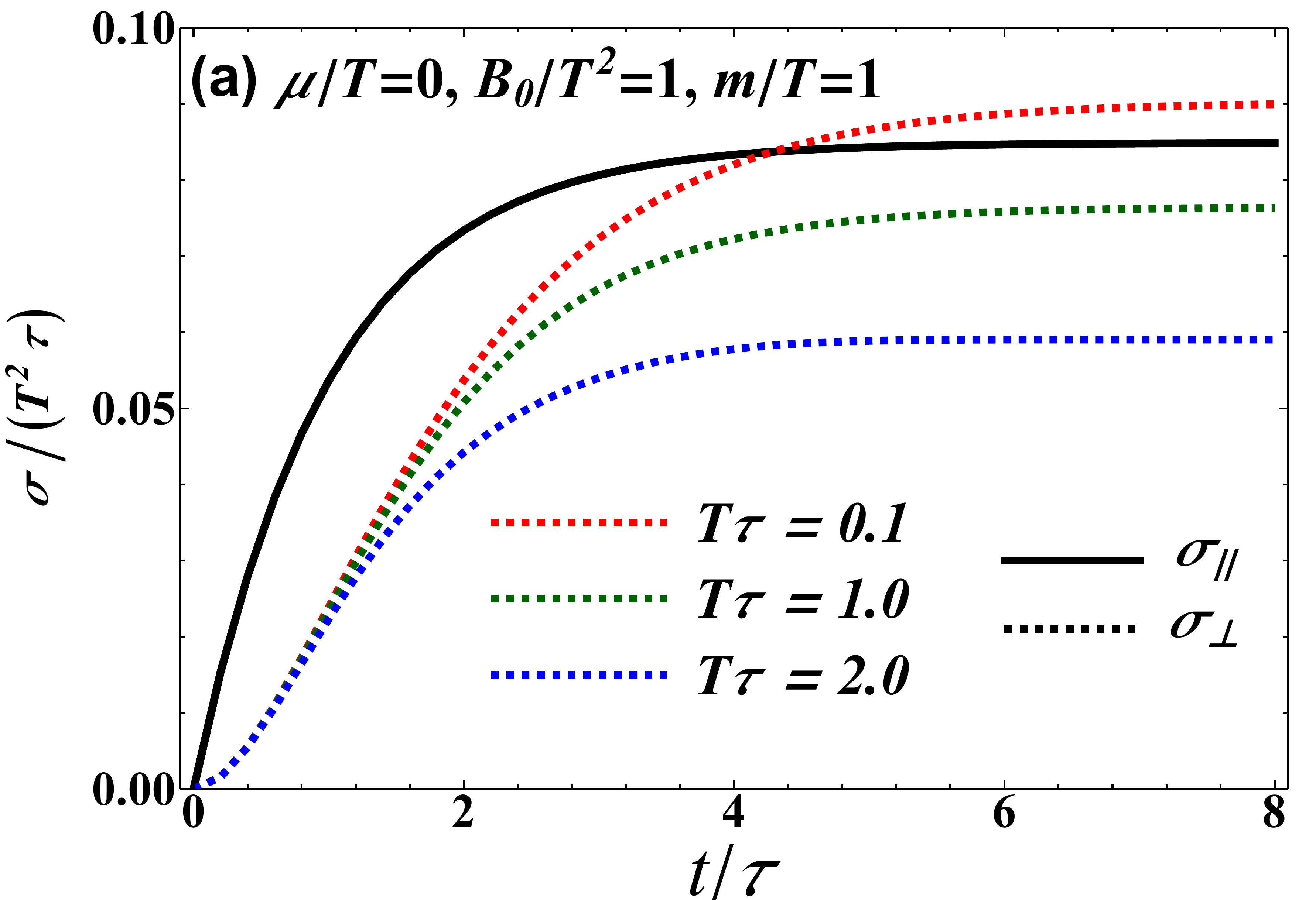}\includegraphics[scale=0.22]{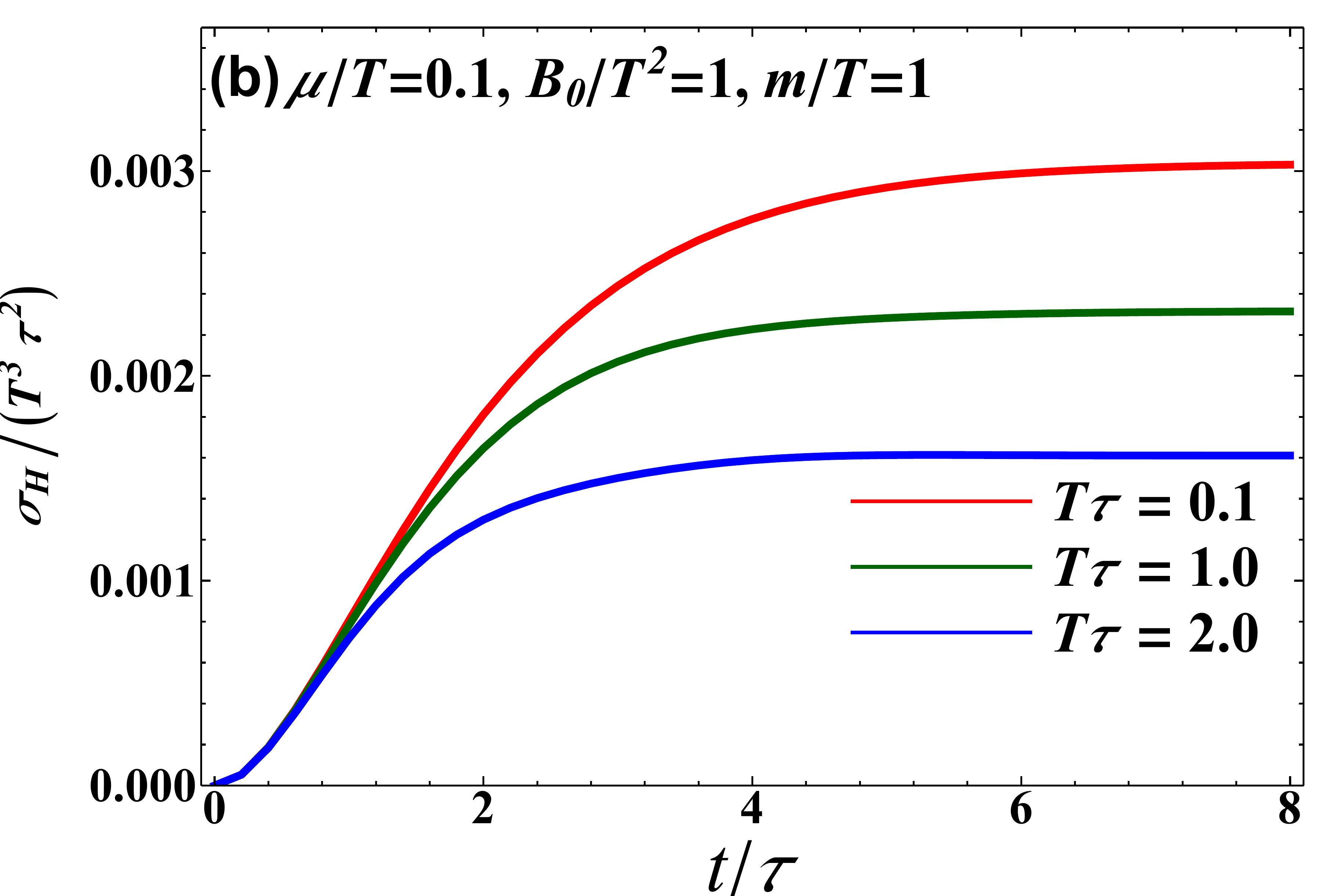}

\caption{\label{fig:The-electric-conductivities} (a) The Ohm conductivities
$\sigma_{\parallel}/(T^{2}\tau)$ and $\sigma_{\perp}/(T^{2}\tau)$
as functions of $t/\tau$ for different values of $T\tau$. The parameters
are set to $\mu/T=0$, $B_{0}/T^{2}=1.0$ and $m/T=1.0$. Note that
$\sigma_{\parallel}/(T^{2}\tau)$ does not depend on $T\tau$ and
is shown in black solid line. The red, green and blue dotted lines
are $\sigma_{\perp}/(T^{2}\tau)$ for $T\tau=$0.1, 1.0 and 2.0, respectively.
(b) The Hall conductivity $\sigma_{H}/(T^{3}\tau^{2})$ as functions
of $t/\tau$ for different values of $T\tau$. The parameters are
set to $\mu/T=0$.1, $B_{0}/T^{2}=1.0$ and $m/T=1.0$.}
\end{figure}

In Fig. \ref{fig:The-electric-conductivities} (a), we show dimensionless
conductivities $\sigma_{\parallel}/(T^{2}\tau)$ and $\sigma_{\perp}/(T^{2}\tau)$
as functions of the time $t/\tau$ for $T\tau=$0.1, 1.0 and 2.0.
As an example, we set $\mu/T=0$, $B_{0}/T^{2}=1$, and $m/T=1$.
We observe that $\sigma_{\parallel}/(T^{2}\tau)$ does not depend
on $T\tau$, as expected, while $\sigma_{\perp}/(T^{2}\tau)$ is larger
for smaller $T\tau$. As $t$ increases, both $\sigma_{\parallel}$
and $\sigma_{\perp}$ increase and eventually reach saturated or asymptotic
values when $t$ is sufficiently large. We have verified that these
behaviors hold for any set of parameters $\{\mu/T,\ B_{0}/T^{2},\ m/T\}$.
In Fig. \ref{fig:The-electric-conductivities} (b), we plot $\sigma_{H}/(T^{3}\tau^{2})$
as functions of $t/\tau$ for $T\tau=$0.1, 1.0 and 2.0. We set $\mu/T=0.1$
because the Hall conductivity vanishes at $\mu=0$. We observe that
$\sigma_{H}/(T^{3}\tau^{2})$ increases with $t/\tau$ and reaches
saturated or asymptotic values when $t$ is large enough. For any
set of $\{\mu/T,\ B_{0}/T^{2},\ m/T\}$, the Hall conductivity is
larger for smaller $T\tau$, which is due to the presence of $B_{0}^{2}\tau^{2}$
in the denominator of Eq. (\ref{eq:Auxiliary-functions}).

\begin{figure}
\includegraphics[scale=0.22]{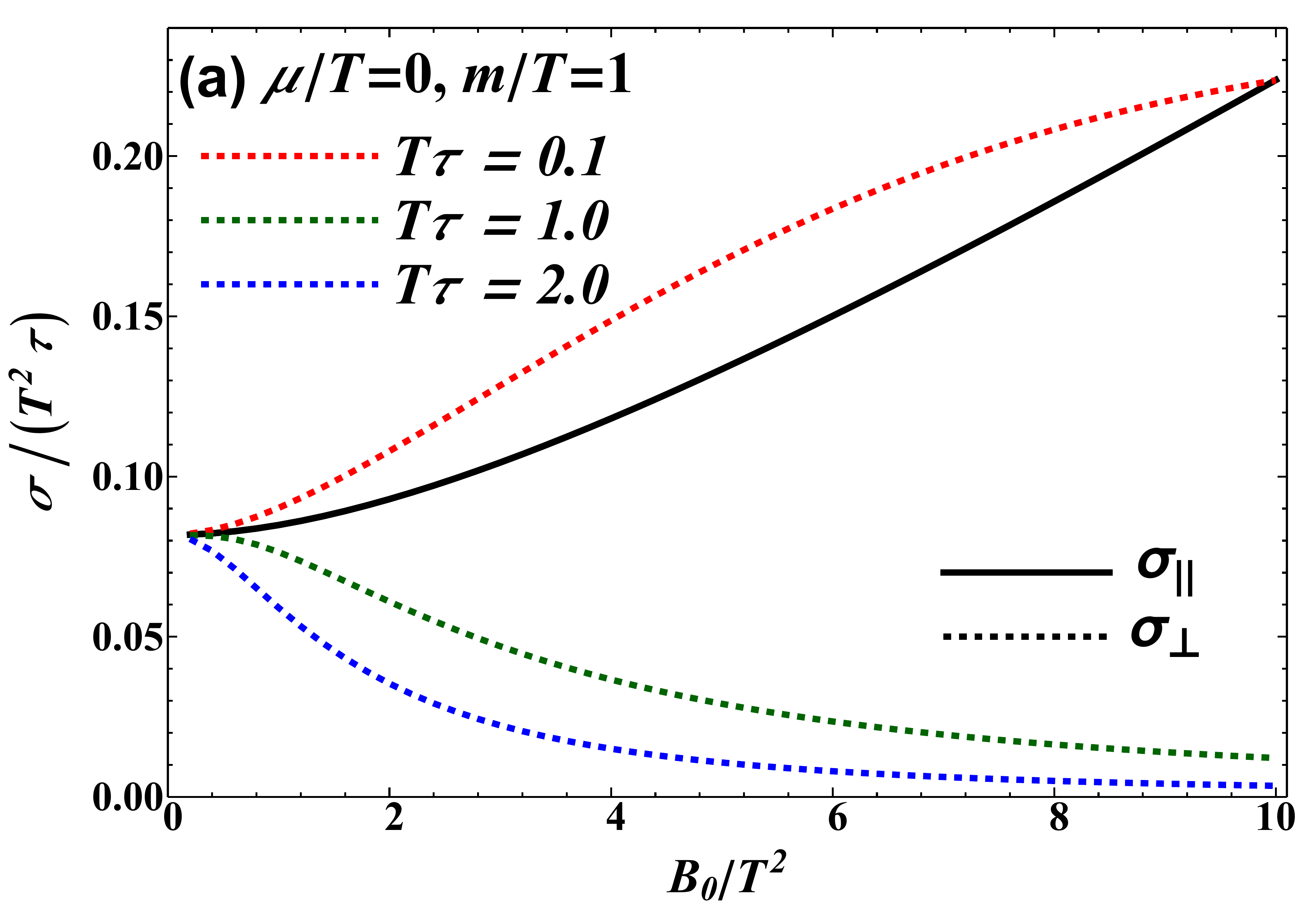}\includegraphics[scale=0.22]{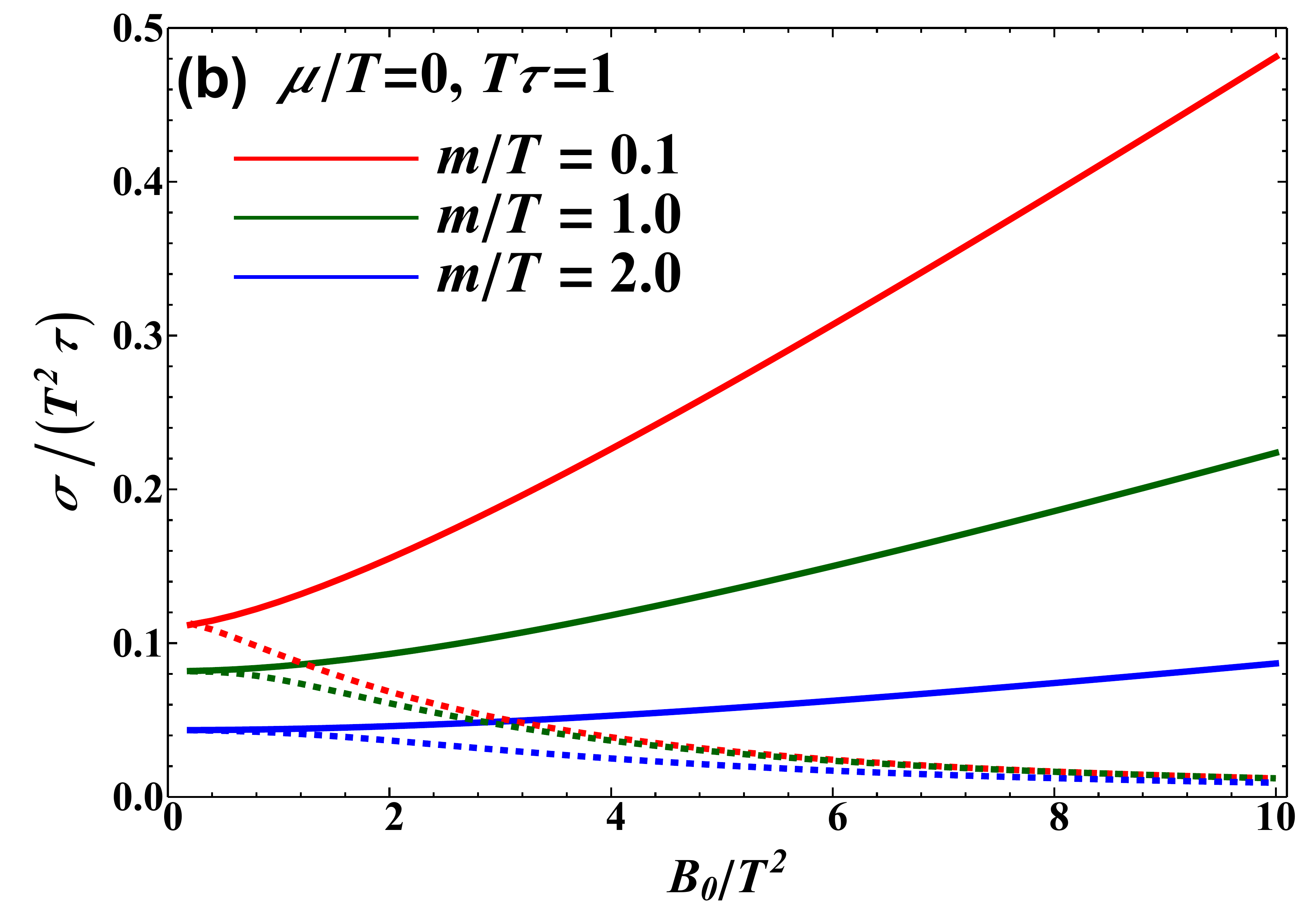}

\caption{\label{fig:stable-electric-1} Asymptotic values at long time limit
for $\sigma_{\parallel}/(T^{2}\tau)$ (solid lines) and $\sigma_{\perp}/(T^{2}\tau)$
(dotted lines) as functions of $B_{0}/T^{2}$. The chemical potential
is set zero. (a) $m/T=1$, black solid line is for $\sigma_{\parallel}/(T^{2}\tau)$
which does not depend on $T\tau$, and the red, green and blue dotted
lines are $\sigma_{\perp}/(T^{2}\tau)$ for $T\tau=$0.1, 1.0 and
2.0, respectively. (b) $T\tau=1$ and $m/T=$0.1 (red), $m/T=$1.0
(green), and $m/T=$2.0 (blue).}
\end{figure}

\begin{figure}
\includegraphics[scale=0.22]{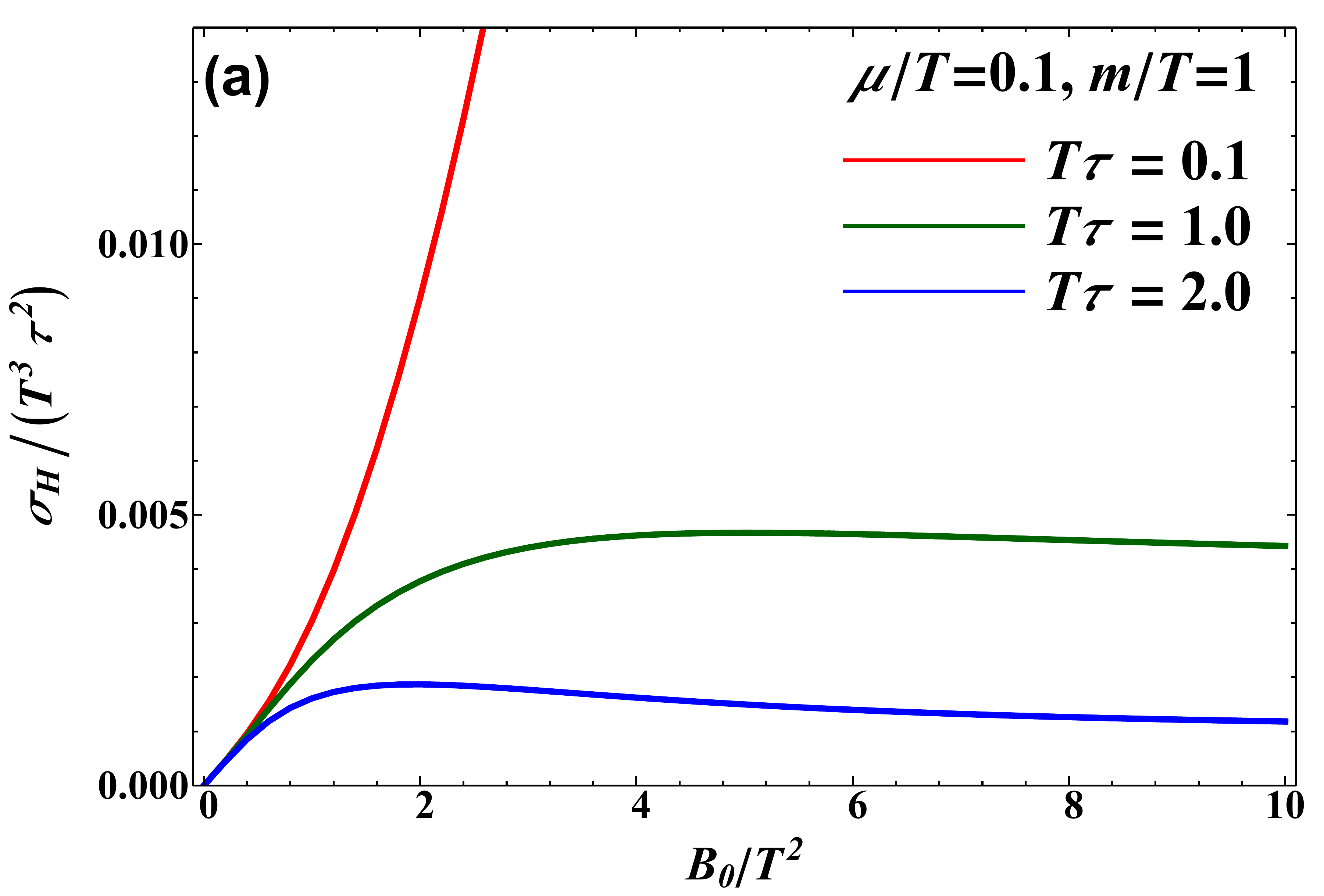}\includegraphics[scale=0.22]{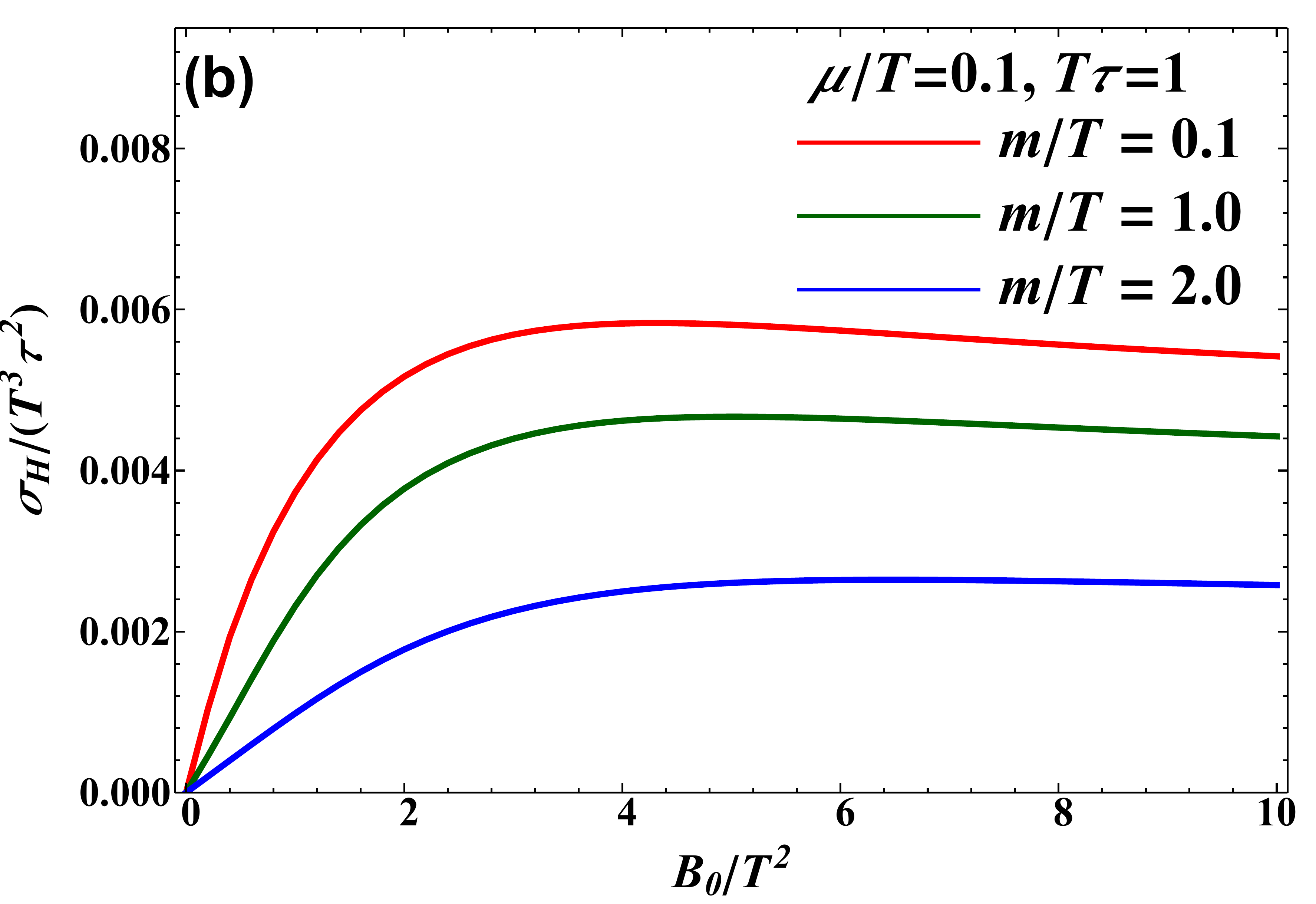}

\caption{The long-time asymptotic values of $\sigma_{H}/(T^{3}\tau^{2})$ (\ref{eq:SigmaHallStable})
as functions of $B_{0}/T^{2}$ with different set of parameter values.
(a) $\mu/T=$0.1, $m/T=1$, $T\tau=$0.1 (red), 1.0 (green), 2.0 (blue);
(b) $\mu/T=0.1$, $T\tau=1.0$, $m/T=$0.1 (red), 1.0 (green), 2.0
(blue). \label{fig:The-stable-electric 2}}
\end{figure}

The long-time asymptotic values of $\sigma_{\parallel}$, $\sigma_{\perp}$,
and $\sigma_{H}$ are presented as functions of $B_{0}/T^{2}$ in
Figs. \ref{fig:stable-electric-1} and \ref{fig:The-stable-electric 2}.
The results for $\sigma_{\parallel}/(T^{2}\tau)$ and $\sigma_{\perp}/(T^{2}\tau)$
are shown in solid and dotted lines respectively. In Fig. \ref{fig:stable-electric-1}(a),
we set $\mu/T=0$, $m/T=1$, the black solid line represents $\sigma_{\parallel}/(T^{2}\tau)$
which does not depend on $T\tau$, as can be seen in Eq. (\ref{eq:SigmaParaStable}),
and the red, green and blue dotted lines represent $\sigma_{\perp}/(T^{2}\tau)$
for $T\tau=$0.1, 1.0 and 2.0 respectively. As $B_{0}/T^{2}\approx0$,
we observe that $\sigma_{\parallel}$ and $\sigma_{\perp}$ are equal.
At finite values of $B_{0}/T^{2}$, there is a difference between
$\sigma_{\parallel}$ and $\sigma_{\perp}$. We find that $\sigma_{\perp}/(T^{2}\tau)$
increases with $B_{0}/T^{2}$ for small $T\tau$, while it decreases
with $B_{0}/T^{2}$ and finally approaches to zero at $T\tau\rightarrow\infty$.
Figure \ref{fig:stable-electric-1}(b) shows the particle mass effects.
We set $m/T=$0.1, 1.0 and 2.0 represented by red, green and blue
lines respectively. For all three cases, we find that $\sigma_{\parallel}/(T^{2}\tau)$
increases while $\sigma_{\perp}/(T^{2}\tau)$ decreases with $B_{0}/T^{2}$.
Such results are qualitatively consistent with those by the lattice
QCD calculation \citep{Astrakhantsev:2019zkr} and the holographic
model \citep{Fukushima:2021got}.


In Fig. \ref{fig:The-stable-electric 2}, we show the long-time asymptotic
values of $\sigma_{H}/(T^{3}\tau^{2})$ in Eq. (\ref{eq:SigmaHallStable})
as functions of $B_{0}/T^{2}$. In Fig. \ref{fig:The-stable-electric 2}(a),
we set $\mu/T=$0.1, $m/T=1$, and $T\tau=$0.1, 1.0, 2.0. In the
weak-field limit we have $\sigma_{H}\propto\tau^{2}$, and then $\sigma_{H}/(T^{3}\tau^{2})$
is independent of $T\tau$. The behaviors of $\sigma_{H}/(T^{3}\tau^{2})$
versus $B_{0}/T^{2}$ are sensitive to the values of $T\tau$. When
$T\tau=2.0$, $\sigma_{H}/(T^{3}\tau^{2})$ increases and then decreases
with $B_{0}/T^{2}$. When $T\tau=1.0$, $\sigma_{H}/(T^{3}\tau^{2})$
increases and then approaches a constant value with $B_{0}/T^{2}$.
When $T\tau=0.1$, $\sigma_{H}/(T^{3}\tau^{2})$ always increases
with $B_{0}/T^{2}$. Figure \ref{fig:The-stable-electric 2}(b) shows
$\sigma_{H}/(T^{3}\tau^{2})$ as functions of $B_{0}/T^{2}$ at $\mu/T=0.1$,
$T\tau=1.0$, and $m/T=$0.1, 1.0, 2.0. For all cases of Fig. \ref{fig:The-stable-electric 2}(a,b),
$\sigma_{H}/(T^{3}\tau^{2})$ is proportional to $B_{0}/T^{2}$ in
the weak field limit. We also find that a large particle mass gives
a small $\sigma_{H}/(T^{3}\tau^{2})$.


\section{Axial-vector current induced by perturbation fields\label{sec:Axial-vector-current-induced}}

In this section, we solve the Boltzmann equation (\ref{eq:Boltzmann-2})
for $\delta\mathcal{A}^{\mu}$ and calculate the axial-current defined
as 
\begin{equation}
J_{A}^{\mu}(x)\equiv\int d^{4}p\,\delta\mathcal{A}^{\mu}(x,p).
\end{equation}
Similar to the decomposition for $\delta\mathcal{V}^{\mu}$ in Eq.
(\ref{eq:decomposing-deltaVmu}) on the bases $e_{i}^{\mu}\equiv\{u^{\mu},b^{\mu},p_{\perp}^{\mu},\epsilon^{\mu\nu\rho\sigma}p_{\nu}^{\perp}u_{\rho}b_{\sigma}\}$,
we can also decompose $\delta\mathcal{A}^{\mu}$ as 
\begin{equation}
\delta\mathcal{A}^{\mu}=u_{\mu}\delta\mathcal{A}_{1}-b_{\mu}\delta\mathcal{A}_{2}-\frac{1}{p_{T}^{2}}\left[p_{\perp}^{\mu}\delta\mathcal{A}_{3}^{\perp}+\epsilon^{\mu\nu\rho\sigma}p_{\nu}^{\perp}u_{\rho}b_{\sigma}\delta\mathcal{A}_{4}^{\perp}\right].\label{eq:decomposing deltaAmu}
\end{equation}
The dynamical equation for each component reads 
\begin{equation}
\left[(u\cdot p)\left(\frac{d}{dt}+\frac{1}{\tau}\right)-B_{0}\epsilon^{\nu\rho\alpha\beta}p_{\nu}\partial_{\rho}^{p}u_{\alpha}b_{\beta}\right]\delta\mathcal{A}_{i}=e_{i}^{\mu}\left(f_{\mu\rho}\mathcal{A}_{\text{eq}}^{\rho}+p^{\nu}f_{\nu\rho}\partial_{p}^{\rho}\mathcal{A}_{\mu}^{\text{eq}}\right),
\end{equation}
where the source term on the right-hand-side can be expressed as
\begin{equation}
e_{i}^{\mu}\left(f_{\mu\rho}\mathcal{A}_{\text{eq}}^{\rho}+p^{\nu}f_{\nu\rho}\partial_{p}^{\rho}\mathcal{A}_{\mu}^{\text{eq}}\right)=D_{i}^{(1)}(t,p_{T},u\cdot p,b\cdot p)+p_{\perp}^{\mu}D_{i,\mu}^{(2)}(t,p_{T},u\cdot p,b\cdot p).
\end{equation}
Similar to $J^{\mu}(x)$ in Eq. (\ref{eq:vector-current}), the axial-vector
current is given by 
\begin{eqnarray}
J_{A}^{\mu}(x) & = & u^{\mu}\int d^{4}p\,g_{1}^{(1)}-b^{\mu}\int d^{4}p\,g_{2}^{(1)}\nonumber \\
 &  & +\frac{1}{2}\int d^{4}p\,\left[g_{3}^{(2)\mu}+\epsilon^{\mu\nu\rho\sigma}u_{\rho}b_{\sigma}g_{4,\nu}^{(2)}\right],\label{eq:axial-vector-current}
\end{eqnarray}
where the functions $g_{i}^{(1)}$ and $g_{i,\mu}^{(2)}$ are derived
from $D_{i}^{(1)}$ and $D_{i,\mu}^{(2)}$ in a similar way to Eq.
(\ref{eq:general-solutions}). Using $\mathcal{A}_{\text{eq}}^{\mu}$
in Eq. (\ref{eq:equilibirum-solutions}), we obtain explicit expressions
for $D_{1}^{(1)}$, $D_{2}^{(1)}$, $D_{3}^{(2)\mu}$, and $D_{4}^{(2)\mu}$
as 
\begin{eqnarray}
D_{1}^{(1)} & = & -(b\cdot p)p_{\parallel}^{\nu}f_{\nu\rho}\sum_{n=0}^{\infty}\left[\partial_{p,\parallel}^{\rho}V_{n}(p)\right]\Lambda_{-}^{(n)}(p_{T}),\nonumber \\
D_{2}^{(1)} & = & -(u\cdot p)p_{\parallel}^{\nu}f_{\nu\rho}\sum_{n=0}^{\infty}\left[\partial_{p,\parallel}^{\rho}V_{n}(p)\right]\Lambda_{-}^{(n)}(p_{T}),\nonumber \\
D_{3}^{(2)\mu} & = & \Xi^{\mu\nu}f_{\nu\rho}\left[(u\cdot p)b^{\rho}-(b\cdot p)u^{\rho}\right]\sum_{n=0}^{\infty}V_{n}(p)\Lambda_{-}^{(n)}(p_{T}),\nonumber \\
D_{4}^{(2)\mu} & = & -\frac{1}{2}\Xi_{\lambda}^{\mu}\epsilon^{\lambda\nu\rho\sigma}p_{\nu}^{\parallel}f_{\rho\sigma}\sum_{n=0}^{\infty}V_{n}(p)\Lambda_{-}^{(n)}(p_{T}).
\end{eqnarray}
Then by making replacements $C_{i}^{(1)}\rightarrow D_{i}^{(1)}$,
$C_{i,\mu}^{(2)}\rightarrow D_{i,\mu}^{(2)}$, $f_{i}^{(1)}\rightarrow g_{i}^{(1)}$
and $f_{i,\mu}^{(2)}\rightarrow g_{i,\mu}^{(2)}$ in Eq. (\ref{eq:general-solutions}),
we are able to derive $g_{1}^{(1)}$, $g_{2}^{(1)}$, $g_{3,\mu}^{(2)}$
and $g_{4,\mu}^{(2)}$. Substituting them into Eq. (\ref{eq:axial-vector-current}),
we obtain 
\begin{eqnarray}
J_{A}^{\mu}(x) & = & -u^{\mu}b^{\nu}\frac{B_{0}}{(2\pi)^{2}}\int_{t_{0}}^{t}dt^{\prime}e^{(t^{\prime}-t)/\tau}\delta E_{\nu}\int d(b\cdot p)\nonumber \\
 &  & \times\lim_{\varepsilon\rightarrow\varepsilon_{b\cdot p}^{(0)}}\,\frac{m^{2}}{\varepsilon^{3}}\left[f_{\text{eq}}^{(+)}(\varepsilon)+f_{\text{eq}}^{(-)}(\varepsilon)\right]\nonumber \\
 &  & -\frac{B_{0}}{(2\pi)^{2}}\Xi^{\mu\nu}\int_{t_{0}}^{t}dt^{\prime}e^{(t^{\prime}-t)/\tau}\delta B_{\nu}\int d(b\cdot p)\nonumber \\
 &  & \times\lim_{\varepsilon\rightarrow\varepsilon_{b\cdot p}^{(0)}}\,\frac{1}{\varepsilon}\sin\left[\frac{B_{0}}{\varepsilon}(t^{\prime}-t)\right]\left[f_{\text{eq}}^{(+)}(\varepsilon)-f_{\text{eq}}^{(-)}(\varepsilon)\right]\nonumber \\
 &  & +\frac{1}{(2\pi)^{2}\tau}\epsilon^{\mu\nu\rho\sigma}u_{\rho}b_{\sigma}\int_{t_{0}}^{t}dt^{\prime}e^{(t^{\prime}-t)/\tau}\left[\left(\frac{d}{dt^{\prime}}+\frac{1}{\tau}\right)\delta B_{\nu}\right]\int d(b\cdot p)\nonumber \\
 &  & \times\lim_{\varepsilon\rightarrow\varepsilon_{b\cdot p}^{(0)}}\,\sin\left[\frac{B_{0}}{\varepsilon}(t^{\prime}-t)\right]\left[f_{\text{eq}}^{(+)}(\varepsilon)+f_{\text{eq}}^{(-)}(\varepsilon)\right].
\end{eqnarray}
Assuming that $\delta E^{\mu}$ and $\delta B^{\mu}$ are proportional
to $\theta(t)$ and they maintain constants at $t>0$, we can express
$J_{A}^{\mu}$ as 
\begin{equation}
J_{A}^{\mu}=-u^{\mu}b^{\nu}\delta E_{\nu}\chi_{A}+\Xi^{\mu\nu}\delta B_{\nu}\chi_{\parallel}+\epsilon^{\mu\nu\rho\sigma}u_{\rho}b_{\sigma}\delta B_{\nu}\chi_{\perp},
\end{equation}
where we still used $\delta E_{\nu}$ and $\delta B_{\nu}$ but to
represent the constant magnitudes of the electric and magentic field
pertubation respectively. The conductivity $\chi_{A}$ denotes the
axial charge induced by the longitudinal electric field $b^{\nu}\delta E_{\nu}$,
and conductivities $\chi_{\parallel}$ and $\chi_{\perp}$ correspond
to the axial current induced by $\delta B_{\nu}$ in the direction
$\Xi^{\mu\nu}\delta B_{\nu}$ and $\epsilon^{\mu\nu\rho\sigma}u_{\rho}b_{\sigma}\delta B_{\nu}$,
respectively, which are given by
\begin{eqnarray}
\chi_{A} & = & \frac{B_{0}\mathcal{R}_{1}(t)}{(2\pi)^{2}}\int d(b\cdot p)\lim_{\varepsilon\rightarrow\varepsilon_{b\cdot p}^{(0)}}\frac{m^{2}}{\varepsilon^{3}}\left[f_{\text{eq}}^{(+)}(\varepsilon)+f_{\text{eq}}^{(-)}(\varepsilon)\right],\label{eq:chiA}\\
\chi_{\parallel} & = & -\frac{B_{0}}{(2\pi)^{2}}\int d(b\cdot p)\lim_{\varepsilon\rightarrow\varepsilon_{b\cdot p}^{(0)}}\frac{1}{\varepsilon}\mathcal{R}_{2}(\varepsilon,B_{0},t)\left[f_{\text{eq}}^{(+)}(\varepsilon)-f_{\text{eq}}^{(-)}(\varepsilon)\right],\label{eq:chiParallel}\\
\chi_{\perp} & = & \frac{1}{(2\pi)^{2}\tau}\int d(b\cdot p)\lim_{\varepsilon\rightarrow\varepsilon_{b\cdot p}^{(0)}}\mathcal{R}_{2}(\varepsilon,B_{0},t)\left[f_{\text{eq}}^{(+)}(\varepsilon)+f_{\text{eq}}^{(-)}(\varepsilon)\right],\label{eq:chiPerp}
\end{eqnarray}
where $\varepsilon_{b\cdot p}^{(0)}\equiv\sqrt{m^{2}+(b\cdot p)^{2}}$
is the on-shell energy at the lowest Landau level and $\mathcal{R}_{1}(t)$,
$\mathcal{R}_{2}(\varepsilon,B_{0},t)$ are defined in Eq. (\ref{eq:Auxiliary-functions}).
We note that only the lowest Landau level contributes to these quantities,
because higher Landau levels are degenerated with respect to spin
and thus do not contribute to the axial current.


Similar to $\sigma_{\parallel}$, $\sigma_{\perp}$, and $\sigma_{H}$
in Eqs. (\ref{eq:SigmaParaStable})-(\ref{eq:SigmaHallStable}), the
conductivities $\chi_{A}$, $\chi_{\parallel}$, and $\chi_{\perp}$
also approach to their long-time asymptotic values that are given
by
\begin{eqnarray}
\lim_{t\rightarrow\infty}\chi_{A} & = & \frac{B_{0}\tau}{(2\pi)^{2}}\int d(b\cdot p)\lim_{\varepsilon\rightarrow\varepsilon_{b\cdot p}^{(0)}}\frac{m^{2}}{\varepsilon^{3}}\left[f_{\text{eq}}^{(+)}(\varepsilon)+f_{\text{eq}}^{(-)}(\varepsilon)\right],\label{eq:chiAStable}\\
\lim_{t\rightarrow\infty}\chi_{\parallel} & = & \frac{B_{0}^{2}\tau^{2}}{(2\pi)^{2}}\int d(b\cdot p)\lim_{\varepsilon\rightarrow\varepsilon_{b\cdot p}^{(0)}}\frac{1}{\varepsilon^{2}+B_{0}^{2}\tau^{2}}\left[f_{\text{eq}}^{(+)}(\varepsilon)-f_{\text{eq}}^{(-)}(\varepsilon)\right],\label{eq:chiParallelStable}\\
\lim_{t\rightarrow\infty}\chi_{\perp} & = & -\frac{B_{0}\tau}{(2\pi)^{2}}\int d(b\cdot p)\lim_{\varepsilon\rightarrow\varepsilon_{b\cdot p}^{(0)}}\frac{\varepsilon}{\varepsilon^{2}+B_{0}^{2}\tau^{2}}\left[f_{\text{eq}}^{(+)}(\varepsilon)+f_{\text{eq}}^{(-)}(\varepsilon)\right].\label{eq:chiPerpStable}
\end{eqnarray}
Note that $\chi_{A}$ is exactly the same as the lowest Landau level
contribution to $\sigma_{\parallel}$. This is because particles at
the lowest Landau level always have spin along the direction of $b^{\mu}$,
thus they are right-handed when their momenta are parallel to $b^{\mu}$,
and left-handed when their momenta are anti-parallel to $b^{\mu}$.
Therefore these particles have the same contribution to the longitudinal
vector current and the axial charge.


\section{Conclusion\label{sec:Conclusion}}

We study the linear response of a fermion system to the perturbative
electromagnetic field in a constant magnetic background field based
on the Wigner function in relaxation time approximation. The Wigner
function is separated into an equilibrium and an off-equilibrium part.
The equilibrium part is the solution to the Wigner function in the
constant magnetic background field, which depends on the thermodynamical
variables.

The off-equilibrium part of the vector and axial vector components
induced by the field perturbation can be determined by solving kinetic
or Boltzmann equations. They also satisfy the mass-shell constraints.
In the case that the field perturbation only depends on the proper
time, the kinetic equations can be solved analytically. The vector
current only depends on the perturbative electric field, with the
linear response are described by longitudinal and transverse Ohm conductivities
as well as the Hall conductivity. As an example, we assume that the
perturbative electric field is turned on at the initial proper time
and maintains constant at later time, we numerically calculate these
conductivities. In the long time limit, these conductivities approach
to asymptotic values which depend on the relaxation time, the particle
mass and the background magnetic field. The contributions from all
Landau levels are included and therefore the results in this paper
can be applied to any strength of the magnetic field. The results
for Ohm conductivities qualitatively agree with those from the lattice
QCD \citep{Astrakhantsev:2019zkr} and the holographic model \citep{Fukushima:2021got}.
We also obtain the analytical solution to the axial vector current
as the linear response to the perturbative electric and magnetic fields.

\begin{acknowledgments}
The authors thank Lixin Yang for enlightening discussions. H.H.P,
S.P. and Q.W. are supported by National Natural Science Foundation
of China (NSFC) under Grants No. 12135011, 12075235 and 11890713 (a
subgrant of 11890710). X.L.S. is supported by the Project funded by
China Postdoctoral Science Foundation under grant No. 2021M701369.
\end{acknowledgments}

\bibliographystyle{h-physrev}
\bibliography{Ref}

\end{document}